%% file: empirical-af.tex
\PassOptionsToPackage{usenames,dvipsnames}{xcolor}
\documentclass[modern]{aastex631}
\usepackage{microtype}  % ALWAYS!
\usepackage{amsmath}
\usepackage{amsfonts}
\usepackage{amssymb}
\usepackage{cancel}
\usepackage{booktabs}
\usepackage{graphicx}

\usepackage{enumitem}
\setlist[description]{style=unboxed}

\sloppypar

\graphicspath{{}}
\input{preamble.tex}

\newcommand{\freqzero}{\ensuremath{\Omega_0}}
\newcommand{\mmax}{\ensuremath{M}}
\newcommand{\rz}{\ensuremath{r_z}}
\newcommand{\rzp}{\ensuremath{\tilde{r}_z}}

\newcommand{\thzp}{\ensuremath{\tilde{\theta}_z}}

\newcommand{\meanY}{\ensuremath{\langle Y \rangle}}

\shorttitle{}
\shortauthors{Price-Whelan et al.}

\begin{document}

\title{
    Data-driven Dynamics with Orbital Torus Imaging:  \\
    A Flexible Model of the Vertical Phase Space of the Galaxy
}

\newcommand{\affcca}{
    Center for Computational Astrophysics, Flatiron Institute, \\
    162 Fifth Ave, New York, NY 10010, USA
}
\newcommand{\affnyu}{
    Center for Cosmology and Particle Physics, Department of Physics, New York University, \\
    726 Broadway, New York, NY 10003, USA
}
\newcommand{\affmpia}{
    Max-Planck-Institut fur Astronomie,
    Konigstuhl 17, D-69117 Heidelberg, Germany
}
\newcommand{\affcolumbia}{
    Department of Astronomy, Columbia University,
    550 West 120th Street, New York, NY 10027, USA
}
\newcommand{\affqueens}{
    Department of Physics, Engineering Physics and Astronomy, Queen's University, \\
    Kingston, K7L 3X5, Canada
}
\newcommand{\affmerced}{
    Department of Physics, University of California, Merced, \\
    5200 Lake Road, Merced, CA 95343, USA
}
\newcommand{\affsurrey}{
    School of Mathematics \& Physics, University of Surrey, Guildford, GU2 7XH, UK
}

\author[0000-0003-0872-7098]{Adrian~M.~Price-Whelan}
\affiliation{\affcca}
\email{aprice-whelan@flatironinstitute.org}
\correspondingauthor{Adrian M. Price-Whelan}

\author[0000-0001-8917-1532]{Jason~A.~S.~Hunt}
\affiliation{\affcca}
\affiliation{\affsurrey}

\author[0000-0003-1856-2151]{Danny~Horta}
\affiliation{\affcca}

\author[0000-0001-5636-3108]{Micah~Oeur}
\affiliation{\affmerced}

\author[0000-0003-2866-9403]{David~W.~Hogg}
\affiliation{\affcca}
\affiliation{\affmpia}
\affiliation{\affnyu}

\author[0000-0001-6244-6727]{Kathryn Johnston}
\affiliation{\affcolumbia}

\author[0000-0001-6211-8635]{Lawrence Widrow}
\affiliation{\affqueens}

% \author{Benjamin~Cassese}

% \author{Neige Frankel}

\begin{abstract}\noindent

The vertical kinematics of stars near the Sun can be used to measure the total mass
distribution near the Galactic disk and to study out-of-equilibrium dynamics.
With contemporary stellar surveys, the tracers of vertical dynamics are so numerous and
so well measured that the shapes of underlying orbits are almost directly visible in the
data through element abundances or even stellar density.
These orbits can be used to infer a mass model for the Milky Way, enabling constraints
on the dark matter distribution in the inner galaxy.
Here we present a flexible model for foliating the vertical position--velocity phase
space with orbits, for use in data-driven studies of dynamics.
The vertical acceleration profile in the vicinity of the disk, along with the orbital
actions, angles, and frequencies for individual stars, can all be derived from that
orbit foliation.
We show that this framework --- ``Orbital Torus Imaging'' (OTI) --- is rigorously
justified in the context of dynamical theory, and does a good job of fitting orbits to
simulated stellar abundance data with varying degrees of realism.
OTI (1) does not require a global model for the Milky Way mass distribution, and (2)
does not require detailed modeling of the selection function of the input survey data.
We discuss the approximations and limitations of the OTI framework, which currently
trades dynamical interpretability for flexibility in representing the data in some
regimes, and which also presently separates the vertical and radial dynamics.
We release an open-source tool, \texttt{torusimaging}, to accompany this article.
\end{abstract}

% \keywords{}

\section{Introduction} \label{sec:intro}

The distribution of mass in the Milky Way is key to many of the questions of
astrophysics.
For one, the mass density determines the orbits of its gas, stars, star clusters, and
satellite galaxies, and thus enables interpreting the kinematic snapshot we observe in
terms of dynamical and galactic evolutionary processes
\citep[e.g.,][]{Freeman:2002,Helmi:2020}.
For another, the distribution of mass depends on the properties of the dark matter on
the scale of an individual galaxy \citep[e.g.,][]{Bertone:2005, Buckley:2018}, which in
turn has implications for fundamental physics and the cosmological model.
On Galaxy mass scales (and smaller), effective models for dark matter predict different
density profiles and different populations of substructures
\citep[e.g.,][]{Bullock:2017}.
Precise measurements of the dark matter within the Milky Way and other nearby galaxies
therefore constrain the particle nature of dark matter.
As a last example, the phase-space distribution function of stars encodes, at some
level, the assembly history of the Galaxy \citep[e.g.,][]{Helmi:2018, Belokurov:2018},
and thus the initial conditions of the primordial patch that evolved into our home in
the Universe \citep[e.g.,][]{Peebles:2011}.

Until direct measurements of the Galactic acceleration field become more ubiquitous
\citep{Klioner:2021, Chakrabarti:2021}, our best hope for studying the mass and dark
matter content of the Milky Way comes from modeling stellar kinematics
\citep[e.g.,][]{Oort:1932, Binney:2008, Rix:2013}.
The principal challenge of this problem is that we only observe a \emph{snapshot} of the
kinematics (i.e. position $\bs{x}$ and velocity $\bs{v}$) of stars throughout the Galaxy
at present day.
We do not observe the orbits of stars or even segments of their orbits, which would
enable a more direct measurement of the acceleration field around those orbits.
Instead, we have to rely on statistical mechanics to relate the snapshot of tracer
kinematics we observe to the underlying mass distribution \citep[e.g.,][]{Kuijken:1989a,
Binney:2008, Magorrian:2014}.

Contemporary stellar surveys provide kinematic data for millions to billions of stars
throughout our Galaxy.
These surveys have enabled new views of the Milky Way that extend far beyond the solar
neighborhood, and have opened up new dimensions for studying the stellar populations of
the Milky Way by providing high-quality stellar labels (e.g., element abundances, ages,
masses, etc.) for large samples of stars.
This includes the transformative astrometric, photometric, and spectroscopic data from
the \gaia\ Mission \citep{Gaia:2016, Gaia:2023}, deep, multi-band photometric surveys
such as the Sloan Digital Sky Survey (SDSS; \citealt{York:2000}), Dark Energy Survey
(DES; \citealt{DES:2016}), and high-resolution spectroscopic surveys such as the Apache
Point Observatory Galactic Evolution Experiment (APOGEE; \citealt{APOGEE:2017}).
Many other stellar surveys are currently underway or planned for the near future that
will expand (or are expanding) the spatial volume, number of stars, number of measured
properties, and precision of the available data for Milky Way stars, such as LAMOST
\citep{LAMOST:2022}, GALAH \citep{DeSilva:2015, Buder:2022}, WEAVE \citep{WEAVE:2023},
4MOST \citep{deJong:2019}, SDSS-V \citep{Kollmeier:2017}, and the Rubin observatory LSST
\citep{LSST:2009}.
This wealth of stellar survey data presently available brings an opportunity to make
precise measurements of the detailed structure of matter and dark matter throughout the
Milky Way and learn about its formation history by combining stellar label and kinematic
data.

Near-invariant stellar labels are complementary to measurements of phase-space
dimensions and can serve as invariant tracers that provide important information in
dynamical analyses.
This idea has already been used and exploited in many contexts.
For example, additional stellar labels can be added in to the phase-space distribution
function (DF) explicitly to form an ``extended distribution function'' (eDF) $f(\bs{W},
\bs{Y})$ \citep[e.g.,][]{Sanders:2015, Binney:2023}, where here the vector $\bs{W}$
represents the kinematic information (i.e. positions and velocities, or integrals of
motion) and $\bs{Y}$ represents any stellar labels, like element abundances.
This is a powerful approach because it allows for using the stellar labels to help in
the dynamical inference of the mass distribution, but requires making choices about how
to parameterize the form of the eDF and any correlations between the kinematics and
stellar labels.
Another challenge with this approach is that it requires a model for the selection
function of the survey to accurately infer the DF properties, which is often not modeled
at a high enough precision to handle the model flexibility demanded by present data.

A different approach has been to model and study the conditional distribution $f(\bs{W}
\given \bs{Y})$.
When $\bs{Y}$ represents element abundances, these are called ``mono-abundance
populations'' (\citealt{Bovy:2012,Bovy:2016}).
This approach is useful because it is conditional on the stellar labels and therefore
does not necessarily require explicitly parameterizing the relationship between the
kinematics and stellar labels (e.g., one can bin in the parameters $\bs{Y}$ and study
the kinematic DF in those bins).
However, this still requires parameterizing the form of the DF in the kinematic
dimensions, and using a model for the selection function of the survey.

We previously introduced a third approach that involves modeling the complementary
factorization $f(\bs{Y} \given \bs{W})$, which we call ``Orbital Torus Imaging'' (OTI;
\citealt{PW:2021}).
This approach does not require detailed knowledge of the survey selection function in
terms of kinematic quantities as long as there is no strong joint dependence on
kinematics and stellar labels.
That is, the selection function $S$ should be approximately separable in kinematic
quantities and stellar labels such that $S = S(\bs{W}) \,S(\bs{Y})$.
OTI is therefore more robust to spatial selection effects than the eDF or mono-abundance
population approaches to modeling Milky Way disk kinematics.
However, it does still require parameterizing any relationships between the kinematics
and stellar labels.

In this work, motivated by the need for flexible dynamical inference methods that make
fewer assumptions about the form of the potential and DF, we build off of the OTI
framework to outline an approach that only requires modeling the shapes of orbits in
projections of phase space.
A companion paper uses this modeling framework to measure the acceleration field and
properties of the Milky Way disk without assuming a global model for the Galactic
potential \citep{Horta:2023}.

\section{Review of Vertical Dynamics in an Axisymmetric Disk} \label{sec:dynreview}

Our goal, like many past efforts, is to define a path for measuring the mass
distribution underlying a tracer population given a snapshot of kinematic data for the
tracers.
For this work, we will consider stars as the tracers and the particular case of modeling
the mass distribution around the Milky Way disk, but we note that these ideas are
generalizable to other contexts.
Though our eventual hope is to build a method that can handle time dependence and
disequilibrium, we will start with a set of standard assumptions to simplify the setup
and limit the dimensionality of expressions.
In particular, for now we will assume that the Galaxy is in equilibrium, that the
distribution function (\df) is in steady state, that the system is axisymmetric, and
that orbital motion is separable in cylindrical radius $R$ and vertical position $z$.
That is, we assume that the total gravitational potential $\Phi$ is additively separable
such that
\begin{equation}
    \Phi(R, z) \approx \Phi_R(R) + \Phi_z(z) \quad .
\end{equation}
This is a typical assumption for near-circular orbits in an axisymmetric potential, such
as for stellar populations within the Galactic disk.

Under the assumptions stated above, the orbits in such a system are governed by a
gravitational acceleration field $\bs{a}(R, z)$ that is related to the underlying mass
density $\rho$ through Poisson's equation,
\begin{align}
    \frac{1}{R} \, \pderiv{}{R}\left(R \, a_R\right) + \pderiv{a_z}{z}
        &= - 4\pi \, G \, \rho(R, z)
\end{align}
where $\bs{a} = -\nabla \Phi$ is the acceleration.
If we assume that we will always work in a small annular volume (i.e. a small range of
$R$), and that the circular velocity curve as a function of radius $v_c(R)$ is nearly
flat over the volume considered, then the radial gradient in the left hand side will be
small and can be neglected, leaving only the vertical terms
\begin{equation}
    -\pderiv{a_z}{z}\biggr\rvert_R = 4\pi \, G \, \rho_z(z \,;\, R)
\end{equation}
where the expressions are assumed to be valid at a given radius $R$.
% Based on historical measurements of the vertical mass density distribution of the Milky Way disk, commonly-adopted forms for $\rho_z(z)$ are a single exponential, a double exponential, or a $\textrm{sech}(z)^2$ profile.
Very near the galactic midplane, the mass density distribution is approximately
constant,
\begin{equation}
    \rho(z\approx 0) \approx \rho_0
        \rightarrow a_z(z) \approx -4\pi \, G \, \rho_0 \, z
\end{equation}
so the shapes of orbits in the vertical phase space are close to ellipses with an aspect
ratio set by the asymptotic midplane vertical frequency,
\begin{equation}
    \freqzero^2 = 4\pi\,G \, \rho_0 \quad . \label{eq:freqzero}
\end{equation}
Orbits that stray further from the midplane will feel an anharmonic potential such that
the vertical frequency decreases for orbits that reach successively higher maximum
heights, $z_{\textrm{max}}$.
See \citet{Read:2014} for a review of mass-modeling methods that use the vertical
kinematics of stars to infer the vertical density or surface mass density structure of
the Milky Way \citep[see also, e.g.,][for recent work on the vertical dynamics of the
disk]{Widmark:2019, Buch:2019, Widmark:2021, Li:2021}.

Orbits in generic axisymmetric, equilibrium systems permit three integrals of motion
that are useful for summarizing and labeling the orbits.
For example, the energy $E$, $z$-component of angular momentum $L_z$, and the ``third
integral'' $I_3$ are all conserved quantities for orbits in an axisymmetric disk.
A more useful set of integrals of motion are the orbital actions $\bs{J} = (J_R,
J_\phi, J_z)$ \citep{Binney:2008}.
Actions are independent isolating integrals of motion that are special in that they are
also the momentum coordinates of a set of canonical coordinates known as action--angle
coordinates.
In this coordinate system, the angle variables $\bs{\theta}$ are the conjugate position
coordinates that increase linearly with time with a rate set by the orbital frequencies
$\bs{\Omega}$.

For this work, as mentioned above, we will consider only the vertical phase space of the
Galaxy under the assumption that the radial and vertical motion are separable \citep[see
also, e.g.,][]{Oort:1932, Bahcall:1984,Kuijken:1989b, Kuijken:1991, Holmberg:2000,
Li:2021, Green:2023}.
Orbits in the vertical phase space are then fully summarized by the vertical action
$J_z$, and the phase of a star along its orbit is set by the vertical angle $\theta_z$.
The vertical action is defined as the area that an orbit sweeps out in the vertical
phase space, i.e.,
\begin{equation}
    J_z = \frac{1}{2\pi} \, \oint \dd z \, v_z(z) \quad, \label{eq:Jz}
\end{equation}
where this integral is done over a full orbital path, which is a closed curve in the
vertical phase space under our assumptions.
Another important invariant property of orbits is the vertical frequency $\Omega_z$,
which is related to the vertical period of an orbit $T_z = 2\pi / \Omega_z$.
The vertical period can be computed as
\begin{equation}
    T_z = \oint \frac{\dd z}{v_z(z)} \label{eq:Tz}
\end{equation}
where the integral is again done over the orbital path.
In general, the vertical frequency and period of an orbit depends on its action
$\Omega_z = \Omega_z(J_z)$.

\begin{figure*}[t!]
\begin{center}
\includegraphics[width=1\textwidth]{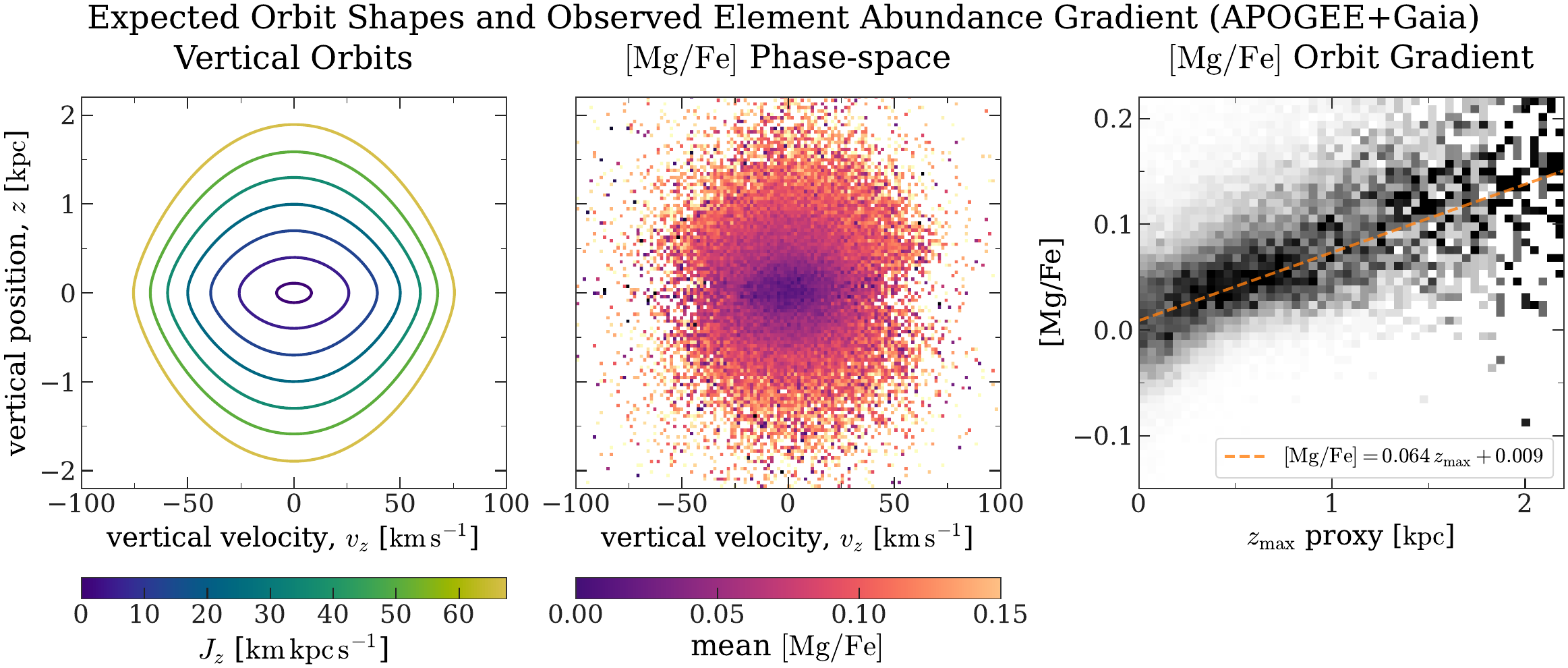}
\end{center}
\caption{%
A demonstration that stellar labels correlate with stellar orbits in the vertical phase
space of the Milky Way disk.
\textbf{Left panel:} Seven orbits computed in a Milky Way mass model (see
Appendix~\ref{sec:appendix-potential}) and shown in the vertical phase space, $(z,
v_z)$.
The orbits have equally-spaced values of the maximum height above the midplane each
orbit reaches, \zmax.
The vertical action, $J_z$, of these orbits scales approximately like $J_z \propto
\zmax^2$, and all orbits have $J_R=0$ and $J_\phi\approx 1900~\kms~\kpc$.
\textbf{Middle panel:} The mean \abun{Mg}{Fe} abundance for stars in the low-$\alpha$
sequence and with angular momenta $L_z$ within $\pm 15\%$ of the value for a circular
orbit at the Sun's position ($L_{z, c} \approx 1900$).
The mean \abun{Mg}{Fe} abundance is shown in small bins of the vertical position, $z$,
and velocity, $v_z$.
The mean abundance is systematically different for stars with low vertical action $J_z$
(i.e. near $(z, v_z) \sim (0, 0)$) as compared to stars with larger vertical action.
\textbf{Right panel:} The column-normalized density of stars in \abun{Mg}{Fe} as a
function of an observable proxy for $\zmax$.
We estimate a proxy for \zmax\ for the data \emph{without using a potential model} by
selecting only stars with $|v_z| < 10~\kms$ (i.e. stars that are near their vertical
apocenter).
The over-plotted dashed (orange) line shows a linear relation between \abun{Mg}{Fe} and
\zmax.
\label{fig:mgfe-zvz}
}
\end{figure*}

\section{Orbital Torus Imaging} \label{sec:oti}

The orbital actions $\bs{J}$ are useful quantities for summarizing orbits and
constructing dynamical models.
However, they require a global model for the gravitational potential, are expensive to
compute, and are not directly observable.
This makes action-based dynamical modeling methods computationally challenging to apply
to large samples and restrictive in terms of the flexibility of the potential model.
Fortunately, for increasingly large samples of stars in the Milky Way, we have access to
measurements of intrinsic, invariant stellar labels, such as stellar surface element
abundance ratios or stellar ages.
These stellar labels can be used instead to trace out the orbit structure of the Galaxy.
This forms the basis of the method of Orbital Torus Imaging (OTI; \citealt{PW:2021}),
which we describe below.
Here we build on past work by rigorously justifying the method and developing a new
implementation of OTI that leverages modern auto-differentiation tools (e.g., \jax;
\citealt{jax:2018}) to generalize and accelerate the model construction and fitting.

\subsection{Using stellar labels to map orbits}
\label{sec:oti-stat}

% From the shapes of these orbits, we can then infer the underlying acceleration field
% and therefore constrain the mass distribution of the Galaxy.
% We will also assume that, for each star, we have measurements of a set of invariant
% ``labels'' $\bs{F}$, and that those labels are correlated with the orbital properties
% of the tracers $\bs{F}(\bs{w})$, for now assuming the dependence is deterministic.

% In an equilibrium system where we have measured stellar kinematics $\bs{w} = (z, v_z)$
% and invariant labels $\bs{F}$ that appear to correlate with the orbits of stars, the
% labels can only depend on the phase-space coordinates through the vertical action
% $J_z$ or any function of the vertical action and not the vertical angle $\theta_z$.
% This comes from the assumption that the labels are invariant: If they depended on
% orbital angle $\theta_z$, this would imply that the labels change over the course of
% an orbit.

% We also then show how, with an OTI model of the orbit structure, this method can be used
% to compute empirical actions, angles, and frequencies that do not require assuming a
% global potential model.

Stars have a number of observable properties that are approximately invariant over their
lifetimes, or at least over galactic orbital timescales.
For example, the surface element abundance ratios, birth time (as a time-invariant proxy
of age), and stellar mass, to name a few.
We represent these quantities with the vector $\bs{Y}$, which we refer to as the
``stellar labels''.
In the Milky Way, these quantities are observed to correlate with the orbital
properties of stellar populations.
For example, there is a long known correlation between stellar age and velocity
dispersion \citep{Spitzer:1951, Sharma:2021}, the kinematic properties of mono-abundance
stellar populations depend on the value of the abundance \citep{Bovy:2016, Yu:2021,
Lian:2022}, and there are known gradients in element abundances and Galactic position
\citep[e.g.,][]{Shaver:1983, Maciel:1999, Eilers:2022, Lian:2023}.
Figure~\ref{fig:mgfe-zvz} shows an example of such a correlation in the vertical
kinematics of stars in the Milky Way.
The left panel of Figure~\ref{fig:mgfe-zvz} shows seven orbits in the vertical phase
space (with $J_R=0$ and $J_\phi \approx 1900~\kms~\kpc$), with equally-spaced values of
\zmax\ as computed in a Milky Way-like gravitational potential around the solar position
(see Section~\ref{sec:appendix-potential}).
The vertical action scales approximately with $J_z \approx \frac{1}{2} \Omega_z \,
\zmax^2$.
The middle panel shows the mean \abun{Mg}{Fe} abundance (as measured by the \apogee\
surveys, \dr{17}; \citealt{APOGEE:2017, APOGEE:DR17}) for stars in different locations
of the vertical phase space.
This demonstrates that stars with low-$J_z$ orbits (i.e. those that stay near the center
of this phase space) have systematically different element abundances than stars with
large-$J_z$ orbits (i.e. those that stay far from the center of this phase space).

% The top left panel of Figure~\ref{fig:sim-contours} shows 8 orbits with different values
% of $J_z$ (but the same values of $J_R=0$ and a $J_\phi$ similar to the solar value)
% computed in a Milky Way-like gravitational potential (see
% Section~\ref{sec:appendix-potential}).
% The orbits are colored by their vertical action, demonstrating that orbits with larger
% vertical actions cover a larger area in vertical phase space and reach higher maximum
% heights above the midplane $z_{\textrm{max}}$.
% Note that the actions define surfaces that foliate the phase space; in the case of
% vertical dynamics, the 1D curves of constant $J_z$ (i.e. the orbits) can never
% intersect.

Under the assumptions listed in Section~\ref{sec:dynreview}, the correlations between
any intrinsic, time-invariant stellar properties $\bs{Y}$ and the orbital properties of
stars can only depend on the orbital actions $\bs{J}$ and not the angles $\bs{\theta}$,
which are time dependent.
This is useful because it means we can use the stellar labels as constants of motion
themselves to trace out the shapes of orbits in phase space.
In the context of vertical dynamics, the vertical action $J_z$ is, by construction,
constant along an orbit, so any time-invariant function of the vertical action is also
constant along an orbit.

With a suitably large sample of stars, this provides a means to measure the orbital
structure and acceleration field of the Galaxy directly, without having to assume a form
for the potential and without needing to compute the actions.
As long as the gradient in stellar labels as a function of vertical action
$\deriv{Y}{J_z}$ is non-zero (i.e. not flat), we can use the level sets of constant
stellar label $Y$ to ``contour'' the orbits in the vertical phase space $(z,
v_z)$.\footnote{We note that this idea, in the context of using the phase-space density
as a stellar label, was briefly discussed in \citet{Kuijken:1989a} and referred to as
``direct contouring'' of the DF. This has been implemented and used previously with
parametric models of the \df\ and potential to study the vertical dynamics
\cite{Li:2021, Li:2023}.}
In reality, the stellar labels are not deterministic functions of the action or of the
vertical phase space: There is a distribution of abundance values at any location of the
phase space $p_Y(Y \given z, v_z)$, as is apparent in Figure~\ref{fig:mgfe-zvz}.
This is expected, given that the processes that induce correlations between the vertical
action, stellar age, and metallicity are inherently stochastic \citep{Nordstrom:2004,
Hayden:2022}.
We will show, however, that moments of the stellar label distribution at different
locations of phase space still enable us to contour the orbits because the distribution
of abundances changes across the vertical phase space.

To see this, we define the (steady-state) extended distribution function (eDF)
$\mathcal{F}$ of the vertical phase space and stellar label $Y$ as
\begin{equation}
    \mathcal{F}(z, v_z, Y) = f(z, v_z) \, p_Y(Y \given z, v_z)  \label{eq:edf}
\end{equation}
where $f(z, v_z)$ is the distribution function (DF) and $p_Y(Y \given z, v_z)$ is the
distribution of stellar labels at each location in phase space.
The DF necessarily satisfies the collisionless Boltzmann equation (CBE; see, e.g.,
\citealt{Binney:2008}), assuming stars are not created or destroyed, which is a
statement of the continuity equation for the phase-space density $f$:
\begin{align}
    \Deriv{f}{t} &= 0 \\
    0 &= \cancel{\pderiv{f}{t}} + \dot{z} \, \pderiv{f}{z} + \dot{v_z} \, \pderiv{f}{v_z} \\
    &= v_z \, \pderiv{f}{z} - \deriv{\Phi}{z} \, \pderiv{f}{v_z} \label{eq:cbe}
\end{align}
where $\dot{x} = \deriv{x}{t}$.
The eDF $\mathcal{F}$ also satisfies a continuity equation: Defining the vector $\bs{w} = (z, v_z, Y)$, we have
\begin{align}
    \pderiv{\mathcal{F}}{t} + \pderiv{}{\bs{w}}
        \cdot (\mathcal{F} \, \dot{\bs{w}}) &= 0 \\
    \pderiv{}{\bs{w}} \cdot (\mathcal{F} \, \dot{\bs{w}}) &=
        \pderiv{}{z} \left(\mathcal{F} \, \dot{z}\right) +
        \pderiv{}{v_z} \left(\mathcal{F} \, \dot{v}_z\right) +
        \pderiv{}{Y} \cancelto{0}{\left(\mathcal{F} \, \dot{Y}\right)} \\
    &= \dot{z} \, \pderiv{\mathcal{F}}{z} + \dot{v}_z \, \pderiv{\mathcal{F}}{v_z}
    \label{eq:F-cbe}
\end{align}
where $\deriv{Y}{t} = 0$ as the stellar labels are assumed to be time invariant.
By expanding the derivatives in Equation~\ref{eq:F-cbe}, this expression simplifies to
\begin{align}
    \dot{z} \, \pderiv{\mathcal{F}}{z} + \dot{v}_z \, \pderiv{\mathcal{F}}{v_z}
    &= p_y \,
    \cancelto{0}{\left(\dot{z} \, \pderiv{f}{z} + \dot{v}_z \, \pderiv{f}{v_z}\right)}
    + f \, \left(\dot{z} \, \pderiv{p_Y}{z} + \dot{v}_z \, \pderiv{p_Y}{v_z}\right)
    \label{eq:cbe-py-tmp} \\
    0 &= \dot{z} \, \pderiv{p_Y}{z} + \dot{v}_z \, \pderiv{p_Y}{v_z} \label{eq:cbe-py}
\end{align}
where the first term in the right-hand side of Equation~\ref{eq:cbe-py-tmp} is set to
zero by the definition of the CBE (Equation~\ref{eq:cbe}).
If we multiply Equation~\ref{eq:cbe-py} by $Y$ and integrate over all $Y$ --- i.e.
compute the first moment of $p_Y$, $\meanY = \int \dd{Y} \, Y p_Y$
--- we get an expression in terms of the mean stellar label,
\begin{align}
    0 &= \pderiv{\meanY}{z} \, \dot{z} +
        \pderiv{\meanY}{v_z} \, \dot{v}_z  \label{eq:cbe-meanY} \quad .
\end{align}
Rearranging Equation~\ref{eq:cbe-meanY}, we can write the vertical acceleration $a_z$ as
\begin{align}
    a_z &= - v_z \, \pderiv{\meanY}{z} \left(\pderiv{\meanY}{v_z}\right)^{-1}
    \label{eq:Y-az}
\end{align}
where the right hand side and partial derivatives are evaluated along a curve of
constant $\meanY$, and we have replaced $\deriv{v_z}{t} = a_z$.
This works for any $Y$ moments of the eDF, $\mathcal{M}_Y = \int \dd{Y} \, M(Y) \,
\mathcal{F}$, where $M(Y)$ is a function of $Y$ that defines the moment of interest.
For simplicity below, as with the expressions above, we only use $M(Y) = Y$ so that we
use the mean stellar label $\meanY$ as our moment of choice.

In the next section, we develop a method for inferring the vertical acceleration profile
$a_z(z)$ that uses this motivation.
In particular, it does not assume a form for the potential and instead parametrizes the
dependence of $\meanY$ on the vertical phase space coordinates, which is sufficient for
inferring the acceleration (Equation~\ref{eq:Y-az}).

% A simple example of this is the case where the stellar label $Y$ is a function of the
% vertical energy $Y=Y(E_z)$ where
% \begin{equation}
%     E_z = \frac{1}{2} v_z^2 + \Phi_z(z)
% \end{equation}
% and $\Phi_z(z)$ is the vertical potential.
% In this case, the partial derivatives in Equation~\ref{eq:Y-az} are:
% \begin{align}
%     \pderiv{Y}{z} &= \deriv{Y}{E_z} \, \deriv{\Phi_z}{z}\\
%     \pderiv{Y}{v_z} &= \deriv{Y}{E_z} \, v_z
% \end{align}
% so that
% \begin{align}
%     a_z &= - v_z \, \deriv{\Phi_z}{z} \, \deriv{Y}{E_z} \, \left(\deriv{Y}{E_z} \, v_z\right)^{-1} \\
%     a_z &= - \deriv{\Phi_z}{z}
% \end{align}
% as expected.

\begin{figure*}[t!]
\begin{center}
\includegraphics[width=0.8\textwidth]{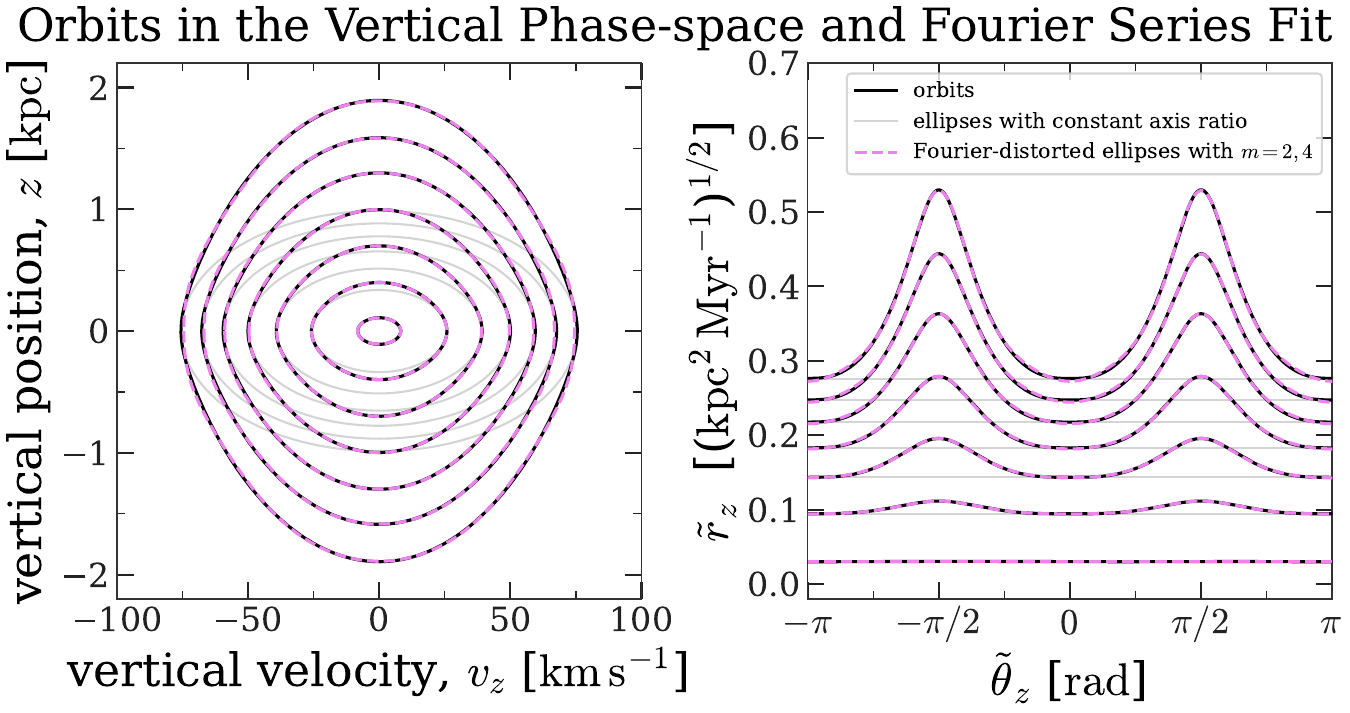}
\end{center}
\caption{%
A demonstration that a low-order Fourier series expansion away from an ellipse can
represent orbit shapes in the vertical phase space of the Milky Way disk.
\textbf{Left panel:} The solid (black) lines show the same orbits as in
Figure~\ref{fig:mgfe-zvz}: These are seven orbits with equally-spaced values of the
maximum height above the midplane each orbit reaches, \zmax, computed in a mass model
for the Milky Way (see Appendix~\ref{sec:appendix-potential}).
The dashed (pink) lines show a Fourier series expansion with $m=\{2, 4\}$ with
parameters ($\rz, e_2, e_4$) determined by fitting to each orbit.
The under-plotted (gray) ellipses show ellipses with the same values of $v_{z,
\textrm{max}}$ as the orbits to emphasize that the orbits have a changing axis ratio
(i.e. frequency) with increased vertical action $J_z$.
\textbf{Right panel:} The same orbits, Fourier series representations, and
constant-frequency ellipses, but now shown in the elliptical radius, \rzp\
(Equation~\ref{eq:rzp}) and angle, \thzp\ (Equation~\ref{eq:thetazp}).
\label{fig:fourier-contours}
}
\end{figure*}

\subsection{A Flexible Model of Stellar Label Moments in Phase Space}
\label{sec:oti-fourier}

In this section, we develop a method to fit the shapes of contours of constant stellar
label moments in the vertical phase space of the Milky Way disk.
An example of the type of data we would like to model is shown in the middle panel of
Figure~\ref{fig:mgfe-zvz}, which shows the mean \abun{Mg}{Fe} abundances of stars in
bins of vertical phase-space coordinates for stars on near-circular orbits in the solar
neighborhood.
As we showed in the previous section (Section~\ref{sec:oti-stat}), these contours of
observed mean \abun{Mg}{Fe} abundance (or any moments of stellar labels) correspond to
orbital trajectories, under the assumptions adopted above.
To parametrize the shapes of the contours of constant stellar label, we use a Fourier
expansion away from an elliptical radius in the vertical phase space, as described
below.

As mentioned above (Section~\ref{sec:dynreview}), orbits near the Galactic midplane ($z
\approx 0$) feel an approximately uniform mass density, $\rho(z) \approx \rho_0$.
This implies that the vertical gravitational potential near the midplane will be close
to a harmonic oscillator potential in which orbits have a constant vertical frequency
$\freqzero$ (Equation~\ref{eq:freqzero}).
These orbits therefore trace closed ellipses of constant elliptical radius \rzp\ defined
as
\begin{equation}
    \rzp = \sqrt{z^2 \, \freqzero + v_z^2 \, \freqzero^{-1}} \label{eq:rzp}
\end{equation}
for all values of the corresponding elliptical angle \thzp,
\begin{equation}
    \tan (\thzp) = \left(\frac{z}{v_z}\right) \, \freqzero \quad .
    \label{eq:thetazp}
\end{equation}
For orbits near the midplane (i.e. with small $J_z$), $\rzp \approx \sqrt{2\,J_z}$.
% where $z_0$ and $v_{z,0}$ are the midplane position and velocity, respectively.

Away from the midplane, the shapes of orbits have steadily decreasing frequencies with
increasing maximum height $z_{\textrm{max}}$, which leads to, to first order, a changing
aspect ratio of the contours.
For example, the orbits in the left panels of Figure~\ref{fig:mgfe-zvz} and
Figure~\ref{fig:fourier-contours} start as horizontal ellipses for the lowest values of
\zmax\ and transition to vertical ellipses for the largest values of \zmax.
For contours with sufficiently large $z_{\textrm{max}}$ values (i.e. orbits that reach
many scale heights above the disk), the decreasing density of the disk and the influence
of the dark matter halo's potential distorts the orbit shapes, and the contours of the
DF are no longer well-approximated by ellipses (e.g., the orbits with $\zmax \gtrsim
1~\kpc$ in Figure~\ref{fig:fourier-contours}).
Orbits with $z_{\textrm{max}} \gtrsim 1~\kpc$ deform from ellipses to more
``diamond-like'' shapes that are ``pinched'' near the midplane ($z=0$).
This shape arises because the orbits feel a nearly uniform density at their vertical
apocenter (i.e. near $v_z = 0$), but then feel a rapidly changing potential as they
cross the midplane (i.e. near $z=0$) with large vertical velocities.

To represent more general contour shapes (i.e. beyond ellipses), we use a low-order
Fourier expansion in the elliptical angle \thzp\ to distort the elliptical radius
\rzp\ (Equation~\ref{eq:rzp}) into the distorted elliptical radius \rz, defined as
\begin{equation}
    \rz = \rzp \, \left[1 + \sum_m \epsilon_m(\rzp) \, \cos{\left(m\,\thzp\right)}\right] \label{eq:rz}
\end{equation}
where, for the vertical kinematics, we only consider even $m$ values to preserve the
symmetry of the contour shapes.
The functions $\epsilon_m(\rzp)$ describe the radius-dependent amplitude of the Fourier
terms, which we require to be zero at $\rzp=0$, and the values are typically $\ll 1$
for the orbits we consider (i.e. with $\zmax \lesssim 2~\kpc$ or so).
This expansion is motivated by the fact that we expect to need just a few Fourier terms:
an $m=2$ distortion with an amplitude that varies with \rzp\ will change the aspect
ratio of the elliptical contours as a function of \rzp, which acts like a changing
frequency as a function of $z_{\textrm{max}}$.
An $m=4$ distortion can ``pinch'' the contour shapes to make then more diamond-like,
which mimics the effect of an orbit feeling the halo potential.
Figure~\ref{fig:fourier-contours} shows the same orbits as in the left panel of
Figure~\ref{fig:mgfe-zvz} (solid black lines) along with a low-order Fourier expansion
fit to these orbits (dashed pink lines), showing that even with just $m=\{2, 4\}$
Fourier terms, we can accurately model the shapes of the orbits.
The left panel of Figure~\ref{fig:fourier-contours} shows the orbits and Fourier fits in
the vertical phase space, and the right panel shows the same orbits and Fourier fits in
the elliptical radius and angle coordinates, \rzp\ and \thzp.

When fitting stellar label data, we assume that the contours of constant stellar label
moments are only functions of the distorted radius, \rz\ (Equation~\ref{eq:rz}). This
then enforces that \rz\ is constant along a contour of constant label value and,
therefore, nearly constant along a vertical orbit.
We then need to specify a functional form for the dependence of the mean stellar label
on \rz, $\meanY(\rz)$, which we assume is a smooth function of \rz\ but is otherwise
customizable (we adopt choices below in the demonstrations in
Section~\ref{sec:applications-sim}.
The model components that must be specified in order to fit this model given a sample of
stellar phase-space positions and labels are: the Fourier distortion amplitude functions
$\epsilon_m(\rzp)$ for all $m$ orders up to the specified \mmax, the functional form and
any parameters of the stellar label function $\meanY(\rz)$, and the asymptotic midplane
orbital frequency \freqzero.
As we will be fitting this model to data where we do not know the true zero-point
position or velocity of the midplane, we additionally include parameters $z_0$ and
$v_{z,0}$ to represent these zero-point values --- we then replace $z$ and $v_z$ in
Equations~\ref{eq:rzp}--\ref{eq:thetazp} with $z-z_0$ and $v_z - v_{z, 0}$.

As the label function only depends on the distorted radius \rz, we can compute the
vertical acceleration from a fitted OTI model directly from the dependence of \rz\ on
the phase space coordinates.
That is, from Equation~\ref{eq:Y-az} and applying the chain rule, we have
\begin{equation}
    a_z = - v_z \, \pderiv{r_z}{z} \left(\pderiv{r_z}{v_z}\right)^{-1} \quad .
    \label{eq:az-rz}
\end{equation}
From the definition of \rz\ and after some manipulation (see
Appendix~\ref{sec:appendix-az}), we find that
\begin{equation}
    a_z(z) = - \freqzero^2 \, z \,
    \frac{\left[
        1 + \sum_{m}^{{2, \dots, \mmax}} (-1)^{m/2} \,
            \left(e_m(\rzp) + \rzp \, \deriv{e_m}{\rzp}\right)
    \right]_{\rzp = \sqrt{\freqzero}\,z}}{\left[
        1 + \sum_{m}^{{2, \dots, \mmax}} (-1)^{m/2} \,
            \left(e_m(\rzp) \, (1 - m^2) + \rzp \, \deriv{e_m}{\rzp}\right)
    \right]_{\rzp = \sqrt{\freqzero}\,z}} \label{eq:az-ugly}
\end{equation}
where $\mmax$ is the maximum (even) order of the Fourier expansion of the distorted
radius, and the expression is evaluated at $\rzp = \sqrt{\freqzero}\,z$.
Note that when $e_m = 0$ for all $m$ (i.e. when there is no distortion of the
elliptical radius), this reduces to the expected expression for the simple harmonic
oscillator.

We obtain the expression above for $a_z(z)$ (Equation~\ref{eq:az-ugly}) by taking the
limit $\thzp \rightarrow \pi/2$, which, in our definition, is equivalent to the limit
$v_z \rightarrow 0$.
However, because our model parametrizes the shapes of orbits without imposing
physicality, our inferred force law can end up depending on velocity.
This is unphysical, but is a tradeoff we make to allow for a flexible model of the phase
space --- we discuss this point further in the Discussion
(Section~\ref{sec:disc-tradeoff}).
Another potential pathology of our model setup is that, for some settings of the
functions $e_m(\rzp)$, the orbits can cross (i.e. the contours of constant label can
intersect).
In a truly separable system, this would also lead to unphysical interpretations of the
force field inferred from the model, as the orbits should foliate the phase space.
We do not find this to be a problem in practice with sufficient bounds set on the
Fourier distortion functions to keep their amplitudes small.

We have implemented this new Orbital Torus Imaging (OTI) framework in \python\ using
\jax\ \citep{jax:2018} to accelerate the model evaluation and to use automatic
differentiation to compute the gradients of the model with respect to the (many) model
parameters.
We release a software package, \texttt{torusimaging}
\citep{torusimaging:zenodo}\footnote{\url{https://github.com/adrn/torusimaging}}, along
with this Article that contains our implementation.
This framework is therefore very general and our implementation accepts any functional
forms for the model components.

\subsection{Computing empirical actions and angles}
\label{sec:empirical-aaf}

The distorted elliptical radius \rz\ will be close to constant along orbits in the
vertical phase space.
The orbital action and frequency should therefore be monotonic and smooth functions of
\rz, i.e. $J_z = J_z(\rz)$ and $\Omega_z = \Omega_z(\rz)$.
We can therefore use a fitted model of the contours of constant stellar label to compute
empirical values of the action, conjugate angle, and frequency for a given phase-space
position $(z, v_z)^*$.
We compute these dynamical quantities using the integrals in
Equations~\ref{eq:Jz}--\ref{eq:Tz} by integrating over the elliptical angle $\thzp$
along a contour of constant $\rz$:
\begin{align}
    J_z &= \frac{2}{\pi} \, \int_0^{\pi/2} \dd \thzp \, v_z(\thzp)
        \, \left|\frac{\dd z}{\dd \thzp}\right| \label{eq:Jz-ell} \\
    T_z &= 4 \, \int_0^{\pi/2} \frac{\dd \thzp}{v_z(\thzp)}
        \, \left|\frac{\dd z}{\dd \thzp}\right| \label{eq:Tz-ell}
\end{align}
where the function $v_z(\thzp)$ and the Jacobian term $\left|\frac{\dd z}{\dd
\thzp}\right|$ are evaluated along a curve of constant $\rz$ corresponding to the
phase-space coordinates $(z, v_z)^*$.
The conjugate angle variable, $\theta_z$, is computed as the fractional time
\begin{equation}
    \theta_z = \frac{2\pi}{T_z} \, \int_0^{\thzp^*} \frac{\dd \thzp}{v_z(\thzp)}
        \, \left|\frac{\dd z}{\dd \thzp}\right| \label{eq:thz-ell}
\end{equation}
where $\thzp^*$ is the value of $\thzp$ corresponding to the phase-space coordinates
$(z, v_z)^*$.

\section{Applications to Simulated Data} \label{sec:applications-sim}

In this section, we demonstrate the OTI framework described above using a set of
simulated datasets of increasing complexity.
We start by using a one-dimensional simple harmonic oscillator potential
(Section~\ref{sec:sim-sho}), then use a more realistic multi-component (3D) galaxy model
(Section~\ref{sec:sim-qiso}--\ref{sec:sim-qiso-sel}), and finally use the final snapshot
from an $N$-body simulation of a disk galaxy perturbed by an orbiting satellite
(Section~\ref{sec:sim-jason}).

Each demonstration below follows the same general procedure: (1) We bin the data into
pixels of the vertical phase space ($z$--$v_z$) and compute the mean element abundance
in each pixel, (2) we fit the OTI model to the binned data and then use a Markov Chain
Monte Carlo (MCMC) sampler to sample the posterior distribution of the model parameters,
and (3) we compute the vertical acceleration profile and other dynamical quantities from
the fitted and sampled models.
Note that, as input to OTI, we bin the data into pixels of vertical position $z$ and
velocity $v_z$ and compute the mean abundance value in each pixel, but we do not
simulate uncertainties on the phase-space coordinates.
This is not a requirement of OTI: We could also use the individual particle positions
and simultaneously handle the uncertainties on the phase-space coordinates by
parameterizing the full distribution $p_Y(Y \given z, v_z)$ (Equation~\ref{eq:cbe-py}).
However, we have found that near the solar neighborhood with modern \gaia\ data, the
uncertainty on the vertical position and velocity is typically comparable to our adopted
pixel size, so we bin the data and work with moments of $p_Y$ to speed up the evaluation
of the likelihood.
In the example cases below, we set the maximum bin edge in each coordinate as three
times the 90th percentile value of the distributions of $z$ and $v_z$ for each simulated
sample, and use 151 bins for each coordinate.
These choices are arbitrary, but we have verified that the results are not sensitive to
reasonable changes of these values.
In the following subsections, we demonstrate that OTI can recover the true vertical
acceleration profiles and other dynamical quantities for these simulated data sets.

\subsection{Simple Harmonic Oscillator}
\label{sec:sim-sho}

\begin{figure*}[t!]
\begin{center}
\includegraphics[width=\textwidth]{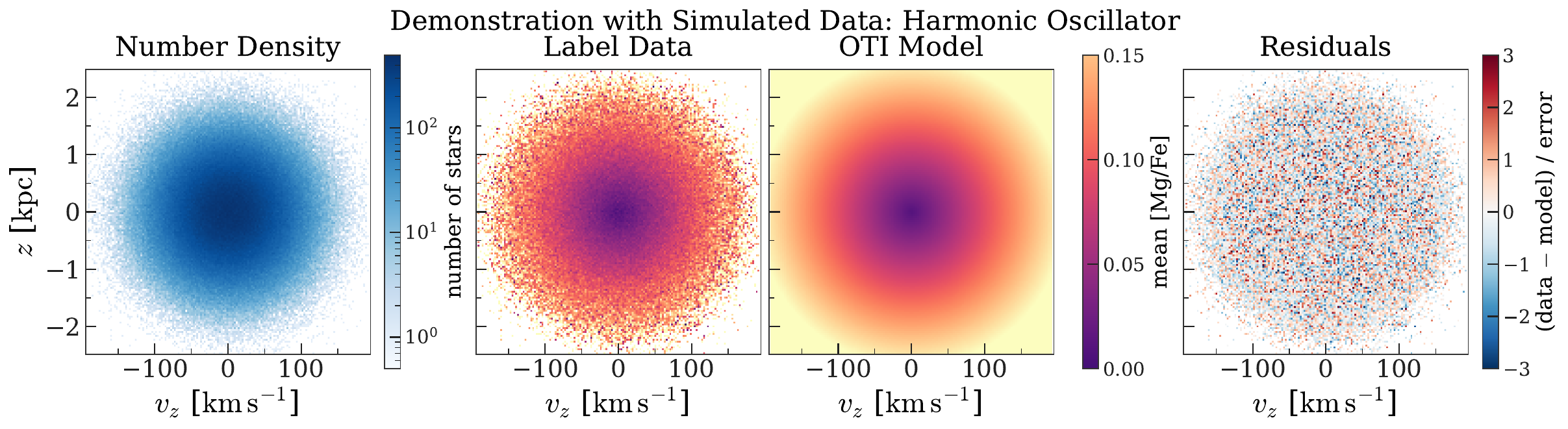}
\end{center}
\caption{%
A demonstration of an optimized OTI model using simulated data in a simple harmonic
oscillator potential.
\textbf{Leftmost panel:} The number density of star particles for a simulated kinematic
dataset in the vertical phase space using an isothermal distribution function embedded in a harmonic oscillator potential.
\textbf{Middle left panel:} The same simulated vertical kinematics now showing the mean
simulated \abun{Mg}{Fe} abundance values, generated assuming a linear relation between
the mean \abun{Mg}{Fe} and $z_{\textrm{max}}$ (see Figure~\ref{fig:mgfe-zvz}).
\textbf{Middle right panel:} The optimized OTI model evaluated on the same grid of
phase-space coordinates as the data (middle left panel).
\textbf{Rightmost panel:} The residuals of the optimized model normalized by the
uncertainty in each pixel (i.e. the simulated data minus the optimized model, evaluated
on the same grid of phase-space coordinates, and normalized).
The small and uniform residuals show that the model accurately represents the mean
abundance trends in this phase space.
\label{fig:sho-data-model}
}
\end{figure*}

As an initial application, we simulate phase-space data in a harmonic oscillator
potential,
\begin{equation}
    \Phi_{z}(z) = \frac{1}{2} \, \omega^2 \, z^2
\end{equation}
by sampling from an isothermal distribution function\footnote{We will use $f(\cdot)$ to
represent a distribution function, \df, that integrates to a number of stars, and
$p(\cdot)$ to represent any normalized probability distribution function (i.e. a \df\
that integrates to 1).}
\begin{align}
    p_z(J_z) &= \frac{1}{\, s_z} e^{-\frac{J_z}{s_z}} \quad ; \quad J_z \in [0, \infty)\\
    s_z &= \sigma_{v_z}^2 / \omega\\
    z = \sqrt{\frac{2 \, J_z}{\omega}} \, \sin\theta_z \quad &; \quad
        v_z = \sqrt{2 \, J_z \, \omega} \, \cos\theta_z
\end{align}
We adopt $\omega = 0.08~\unit{\radian\per\mega\year}$ and $\sigma_{v_z} = 50~\kms$ as
somewhat arbitrary choices meant to approximately match the observed vertical kinematics
of stars near the solar position in the Galactic disk.
We sample $N=2^{18}=\num{262144}$ phase-space positions to use as our simulated data
set, but we note that the precision of our measurements (especially at high $z$) will
depend on the sample size.
We then assign each star a simulated element abundance value to serve as a stellar label
$Y$: We use the relation between \abun{Mg}{Fe} and $z_{\textrm{max}}$ from the \apogee\
data (see the right panel of Figure~\ref{fig:mgfe-zvz}) to assign each star a simulated
value of $Y_{\abun{Mg}{Fe}}$ by drawing from a Gaussian centered on the mean value from
the linear relation with a standard deviation of $0.05~\textrm{dex}$.
That is, our simulated abundance values are drawn from
\begin{equation}
    Y_{\abun{Mg}{Fe}} \sim \mathcal{N}(0.064~\zmax + 0.009, 0.05)
\end{equation}
where $\mathcal{N}(\mu, \sigma)$ is a normal distribution with mean $\mu$ and standard
deviation $\sigma$.
We then additionally simulate measurement errors on the abundance values by drawing an
abundance uncertainty, $\sigma_Y$, for each star particle such that $\log \sigma_Y \sim
\mathcal{U}(-4, 0.5)$, where $\log$ is the natural logarithm and $\mathcal{U}(a, b)$ is
the uniform distribution over the domain $[a, b)$.
Figure~\ref{fig:sho-data-model} (middle left panel) shows the mean simulated
\abun{Mg}{Fe} abundance for stars in bins of vertical phase-space coordinates with the
same color scale as in the middle panel of Figure~\ref{fig:mgfe-zvz} (i.e. the real
\apogee\ data).

We then fit the simulated mean abundance data in pixels of phase-space coordinates,
$\langle Y_{\abun{Mg}{Fe}} \rangle$, using the OTI framework described above
(Section~\ref{sec:oti}).
For this initial test case, we include only an $m=2$ term in our Fourier distortion of
the elliptical radius (Equation~\ref{eq:rz}).
Our input data are the mean label values in each pixel $j$ computed as
\begin{equation}
    \langle Y_{\abun{Mg}{Fe}} \rangle_j =
        \frac{\sum_n^{N_j} \frac{Y_n}{\sigma_{Y,n}^2}}{\sum_n^{N_j} \frac{1}{\sigma_{Y,n}^2}} \label{eq:mean-abun}
\end{equation}
where $Y_n$ and $\sigma_{Y, n}$ are the individual stellar abundance measurements and
their uncertainties for the $N_j$ stars (indexed by $n$) within a given pixel $j$.
To represent this data, we must decide on functional forms for both the label function,
$Y(\rz)$, which will serve as our model prediction for the mean abundance in each pixel,
and for the Fourier coefficient functions, $e_m(\rzp)$, which here is just $e_2(\rzp)$.
For both of these functions, $Y(\rz)$ and $e_2(\rzp)$, we use monotonic quadratic spline
functions defined such that the function is always monotonically increasing or
decreasing, which must be set before optimization.
% (see Appendix~\ref{sec:appendix-spline}).

For the label function, $Y(\rz)$, we assert that the value must monotonically increase
from $\rz = 0$ to the maximum spline knot location by noting that the element abundance
gradient has a positive slope from small values of the phase-space coordinates $(z,
v_z)\sim (0,0)$ to larger values (see the middle left panel of
Figure~\ref{fig:sho-data-model}).
We use $K_Y=8$ spline knot locations $x_k^{(Y)}$ equally spaced in $\rz$ between $0$
and $r_{z, \textrm{max}}\approx 0.7~\rzunit$ with function values at the knot locations
$y_k^{(Y)}$.
This choice is arbitrary: We could here do a form of cross-validation to set the number
of spline knots, but we set the number for simplicity.

For the Fourier coefficient function, $e_2(\rzp)$, we require that the coefficient value
must equal zero at $\rzp = 0$ (i.e. $e_2(0) = 0$), and that the coefficient value
monotonically increases with $\rzp$.
Requiring $e_2(0) = 0$ ensures that the parameter $\freqzero$ corresponds to the
asymptotic orbital frequency as $\rzp \rightarrow 0$ and can therefore be used to
estimate the midplane volume mass density (Equation~\ref{eq:freqzero}).
Assuming that the coefficient value increases with $\rzp$ (so that the orbit shapes
can only become more vertically stretched with increasing $z_{\textrm{max}}$) is
equivalent to assuming that the surface density increases with increasing height $z$.
We use $K_{e_2}=8$ spline knot locations $x_k^{(e_2)}$ with values $y_k^{(e_2)}$ equally
spaced in $\rz^2$ between $0$ and $r_{z, \textrm{max}}$, but we require the knot value
at $\rzp = 0$ to be equal to zero $y_0^{(e_2)}=0$.
% All of the parameters used in this demonstrative model fit and the adopted bounds used
% when optimizing are summarized in Table~\ref{tbl:sho-params}.
When optimizing, we reparameterize to use $\log \freqzero$ and $\log y_k$ for any spline
coefficients as a trick to control the sign of the parameter values.

Our framework is implemented with \jax\ so that our objective function is just-in-time
compiled and we can use automatic differentiation to compute the gradients of the
objective function with respect to the, in this case, 18 model parameters: 8 spline
knots for the label function, 7 spline knots for the $e_2$ function (there are 8 knots,
but the value at zero is fixed to zero, so only 7 free values), 2 parameters for
centering $z_0$ and $v_{z, 0}$, and the asymptotic midplane frequency $\freqzero$.
We use a Gaussian log-likelihood for the mean \abun{Mg}{Fe} abundance data in each pixel
of phase space.
That is, the log likelihood of the data, $\ln \mathcal{L}$, is computed as
\begin{equation}
    \ln \mathcal{L} = \sum_j \ln
        \mathcal{N}(\langle Y_{\abun{Mg}{Fe}} \rangle_j \given Y_j, \sigma_{Y, j})
\end{equation}
where the sum is done over the $j$ pixels, $\langle Y_{\abun{Mg}{Fe}} \rangle_j$ is the
mean abundance value in pixel $j$ (computed as described above;
Equation~\ref{eq:mean-abun}), $Y_j$ is the model predicted mean abundance value in pixel
$j$, and $\sigma_{Y, j}$ is the uncertainty on the mean abundance.
For the uncertainty on the mean abundance in each pixel $\sigma_{Y, j}$, we consider
both the error on the mean (given heteroskedastic measurement errors on the abundances
of each star particle, $\sigma_{Y,i}$),
\begin{equation}
    \sigma_{\mu_Y} = \sqrt{\frac{1}{\sum_i \frac{1}{\sigma^2_{Y,i}}}}
\end{equation}
and the intrinsic scatter of abundances in each pixel.
We assume for this case that we do not know the true intrinsic scatter that we simulated
with in order to emulate working with real data, so we compute the error-deconvolved
intrinsic scatter, $s_y$, of abundance values in the ten pixels with the most number of
star particles.
We take the uncertainty of the mean abundance $\mu_{Y,j}$ in each pixel $j$ to be
\begin{equation}
    \sigma_{\mu_{Y,j}} = \sqrt{\sigma_{\mu_Y, j}^2 + s_Y^2 / N_j}
\end{equation}
where $N_j$ is the number of star particles in pixel $j$ and $\sigma_{\mu_Y, j}$ is the
error on the mean in the same pixel.

As a way to enforce smoothness of the spline functions, we use a Gaussian prior on the
spline function derivatives evaluated at the knot locations with a standard deviation of
0.5 for the label function derivatives, and 0.2 for the Fourier coefficient function
derivatives.
We then optimize the model parameters using the sum of the log likelihood and the log
prior to find the maximum a posteriori (MAP) model parameters, then use this to
initialize a Markov Chain Monte Carlo (MCMC) sampling of the parameters, to assess
uncertainty on the model parameters.

We use a standard L-BFGS-B optimizer \citep{Byrd:1995} to minimize the regularized
negative log-probability as implemented in \texttt{JAXopt} \citep{jaxopt:2021}.
For this simulated data set, the optimization runs in $\sim 10~\unit{\second}$ on a
single CPU.
The middle right panel of Figure~\ref{fig:sho-data-model} shows the MAP model evaluated
over the domain of our simulated data in the left panel, and the right panel shows the
residuals.
The residuals everywhere indicate that the best-fit model matches the data well.

\begin{figure*}[t!]
\begin{center}
\includegraphics[width=\textwidth]{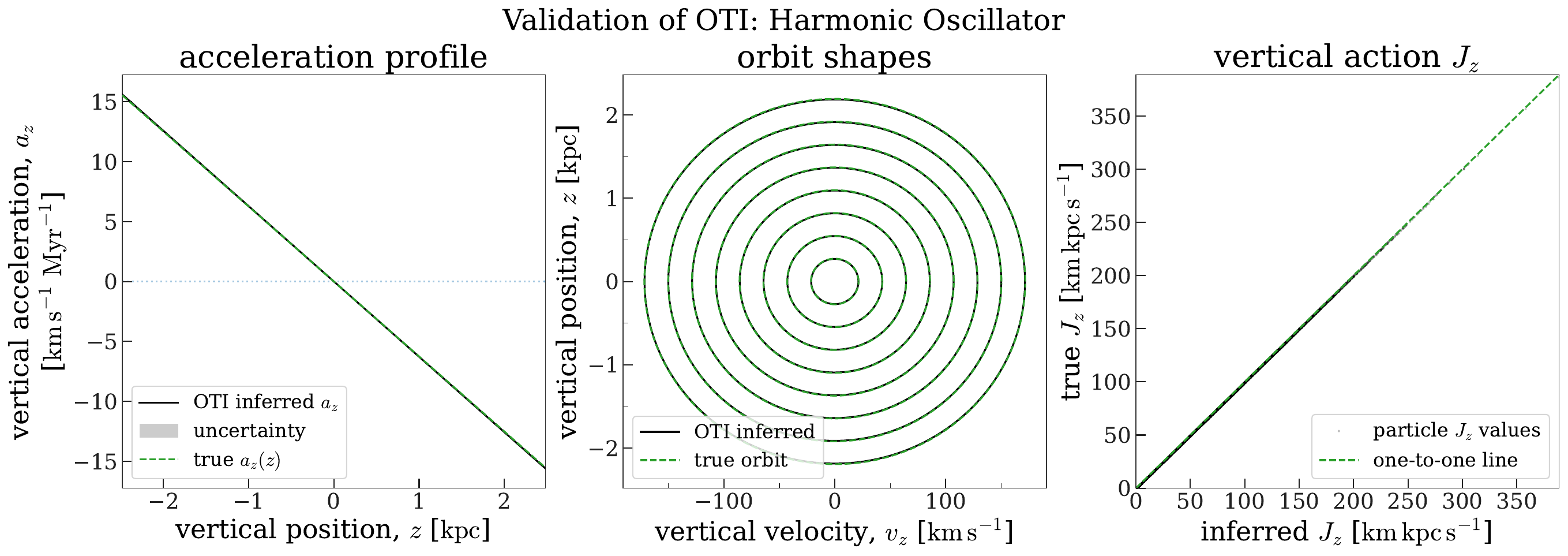}
\end{center}
\caption{%
Validation of the OTI model fit to the simulated simple harmonic oscillator data.
\textbf{Left panel:} The inferred (maximum a posteriori; MAP) vertical acceleration
profile, $a_z(z)$, from the OTI model (solid black line) compared to the true
acceleration profile (dashed green line).
The uncertainty in the acceleration trend is comparable to the width of the line.
Most of the simulated data are within $|z| \lesssim 2~\kpc$ and the inferred
acceleration profile agrees well with the true values within this region.
\textbf{Middle panel:} Eight orbits with equally-spaced values of the vertical velocity
computed in the true potential model (simple harmonic oscillator; shown as dashed, green
lines) and with the OTI model using the MAP parameter values (solid, black lines).
\textbf{Right panel:} A comparison of the true vertical action values (vertical axis)
and our estimates using the MAP OTI model (horizontal axis).
Our action estimates are empirical in that we do not need to assume the form for a
gravitational potential.
The median (fractional) action error is $<0.5\%$ for this sample dataset.
\label{fig:sho-validation}
}
\end{figure*}

We use the optimized parameter values to initialize two MCMC chains using the
Hamiltonian Monte Carlo ``No U-turn Sampler'' (NUTS) implemented in the
\texttt{blackjax} \citep{blackjax2020github} Python package.
We run the chains for 1000 initial steps to allow the NUTS sampler to tune the
hyperparameters of the sampling procedure (e.g., the mass matrix), and then run for an
additional 1000 steps.
The left panel of Figure~\ref{fig:sho-validation} shows the maximum a posteriori (MAP)
vertical acceleration profile, $a_z(z)$ (computed with Equation~\ref{eq:az-ugly}), as
the under-plotted black line, and the over-plotted, dashed (green) line shows the true
vertical acceleration relation, $a_z = -\omega^2 \, z$.
This demonstrates that, for this toy dataset, we are able to recover the true
acceleration profile using OTI.
In the same panel, we also plot a shaded (gray) polygon to represent the acceleration
profile span for the 16th--84th percentile values of the MCMC samples.
However, in this case, the uncertainty region is comparable to the thickness of the
black line.

The middle panel of Figure~\ref{fig:sho-validation} shows eight orbits with
equally-spaced (but arbitrary) values of the vertical velocity at $z=0$: The
under-plotted solid (black) lines show the orbital trajectories inferred using the MAP
parameter values with our OTI model, and the over-plotted dashed (green) lines show the
true orbital trajectories for this toy model, demonstrating agreement.
The right panel of Figure~\ref{fig:sho-validation} shows the true vertical action values
for all simulated particles on the vertical axis (i.e. the samples from the \df\ used to
create the mean abundance data shown in the middle left panel of
Figure~\ref{fig:sho-data-model}) and our estimates for the vertical action of each star
particle as the black point markers, computed using the MAP parameter values.
We compute the actions, frequencies, and angles using
Equations~\ref{eq:Jz-ell}--\ref{eq:thz-ell} evaluated at the phase-space coordinates of
each star particle.
The acceleration error at $z=1~\kpc$ is $< 1\%$ (in fractional error) and the median
action error is $<0.5\%$ for this toy dataset.

For this demonstration, we do not have to define a global model for the gravitational
potential or the \df.
We are able to recover the vertical acceleration profile using a flexible model of the
shapes of orbits as traced by the stellar labels.
The advantage of using very flexible spline functions for the label function and Fourier
distortion functions is that our model is able to fit the input data well while making
fewer strong assumptions.
However, the disadvantage is that the model is not constrained to be physical and can
therefore end up with, e.g., a Fourier distortion function form that violates
physicality in the sense that the implied density could be negative or change
discontinuously.
In the next section, we show how this performs for a more realistic (but still
equilibrium) galaxy model.

\subsection{Axisymmetric Quasi-isothermal Disk}
\label{sec:sim-qiso}

\begin{figure*}[t!]
\begin{center}
\includegraphics[width=\textwidth]{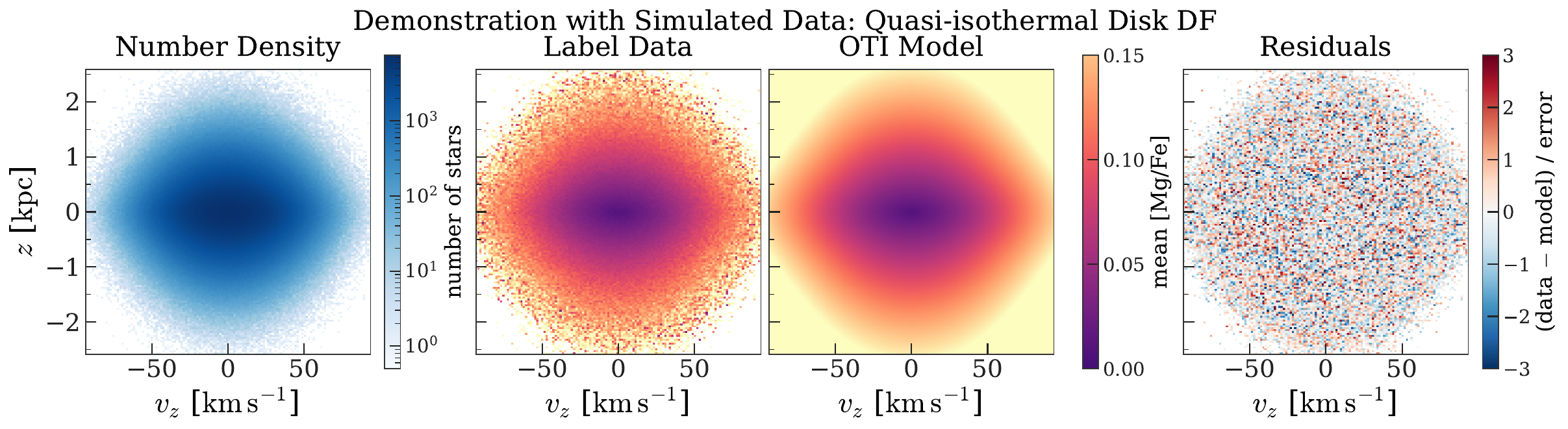}
\end{center}
\caption{%
The same as Figure~\ref{fig:sho-data-model}, but for the simulated quasi-isothermal disk
sample from Section~\ref{sec:sim-qiso}, showing the simulated data set (left and middle
left panels), the optimized OTI model (middle panel), and the residuals (right panel).
% \textbf{Left panel:} Simulated kinematic data in the vertical phase space with an
% isothermal distribution function and a linear relation between \abun{Mg}{Fe} and
% $z_{\textrm{max}}$ (see Figure~\ref{fig:mgfe-zvz}).
% \textbf{Middle panel:} An optimized OTI model evaluated on the same grid of phase-space
% coordinates as the data (left panel).
% \textbf{Right panel:} The residuals of the best-fit model (i.e. the simulated data minus
% the best-fit model evaluated on the same grid of phase-space coordinates).
\label{fig:qiso-data-model}
}
\end{figure*}

As a next application of the OTI framework, we use a more realistic galaxy model to
sample 3D positions and velocities of star particles and show that OTI can still recover
the vertical dynamics.
We use a quasi-isothermal disk distribution function \citep[\df;][]{Binney:2012} with
parameters derived from fitting Milky Way stellar data (we use the ``thin disk''
parameters from Table~3 of \citealt{Sanders:2015}).
We implement this \df\ using the \agama\ package \citep{Vasiliev:2019} and sample
\num{5d8} phase-space coordinates with guiding-center radii $R_G$ between $5 < R_G <
12~\kpc$ embedded in our adopted simple Milky Way potential model (see
Appendix~\ref{sec:appendix-potential}).
We compute the guiding radii using an approximation assuming the circular velocity
curve, $v_c(R)$, is flat, $R_G = L_z / v_c(R)$, where $L_z$ is the angular momentum of
the star particle and $R$ is its instantaneous cylindrical radius.
From this simulated sample, we select only stars near the solar radius, $R_0=8.3~\kpc$,
with $|R-R_0| < 1~\kpc$, and additionally select $|R-R_G| < 1~\kpc$ and $|v_R| <
15~\kms$ as a way of limiting the radial action of the stars so that our assumption of
$R$--$z$ separability is more valid.
We do not use the radial action itself to select stars because we want to mimic the
kinds of selections we could do with real data, where we do not know the true underlying
gravitational potential (which would be needed to compute the actions).
After these selections, we are left with $\approx \num{7d6}$ star particles.
We again compute simulated \abun{Mg}{Fe} abundance values for these stars using the
linear relation between \abun{Mg}{Fe} and $z_{\textrm{max}}$ from the \apogee\ data (see
the right panel of Figure~\ref{fig:mgfe-zvz}) and a scatter of 0.05~dex (see
Section~\ref{sec:sim-sho}), with simulated uncertainties again following the
prescription outlined previously (Section~\ref{sec:sim-sho}).
The left panels of Figure~\ref{fig:qiso-data-model} shows this simulated data set in the
vertical phase space coordinates.

We model the simulated mean abundance data in pixels of phase-space coordinates using
the OTI framework described above (Section~\ref{sec:oti}) and following the same
procedure for optimizing and then MCMC sampling the model as in
Section~\ref{sec:sim-sho}.
%(Appendix~\ref{sec:appendix-spline})
We again use monotonic quadratic spline functions for both the label function and the
Fourier coefficient functions.
For the label function, we use 8 knots equally spaced in \rz\ between $0$ and $r_{z,
\textrm{max}} \approx 0.55~\rzunit$.
However, here we use $m=\{2, 4\}$ Fourier terms with 12 and 4 spline knots,
respectively.
We space the Fourier coefficient spline knots equally in $\rz^2$ between $0$ and
$r_{z, \textrm{max}}$: This places a higher density of knots at lower \rz\ values where
we expect the Fourier coefficient functions to change more rapidly.
We again require that the coefficient function values are zero at $\rzp = 0$ so that
$e_2(0)=e_4(0)=0$, and we assume that the overall sign of the functions are $(+,-)$ for
the $m=\{2, 4\}$ coefficient functions, respectively.

We use the same priors as in Section~\ref{sec:sim-sho} to lightly enforce smoothness of
the derivatives of the spline functions.
As above, we initially use a L-BFGS-B optimizer to minimize the regularized negative
log-probability.
This model has 22 parameters and optimizes in $\sim 30$~seconds on a single CPU.
The middle right panel of Figure~\ref{fig:qiso-data-model} shows the best-fit model
evaluated on the same pixel grid as the simulated data, and the rightmost panel shows
the residuals between the optimized model and the data, normalized by the uncertainty on
the mean abundance in each pixel.
The small residuals indicate that this model does well to fit the mean abundance
variations in the vertical phase space.

\begin{figure*}[t!]
\begin{center}
\includegraphics[width=\textwidth]{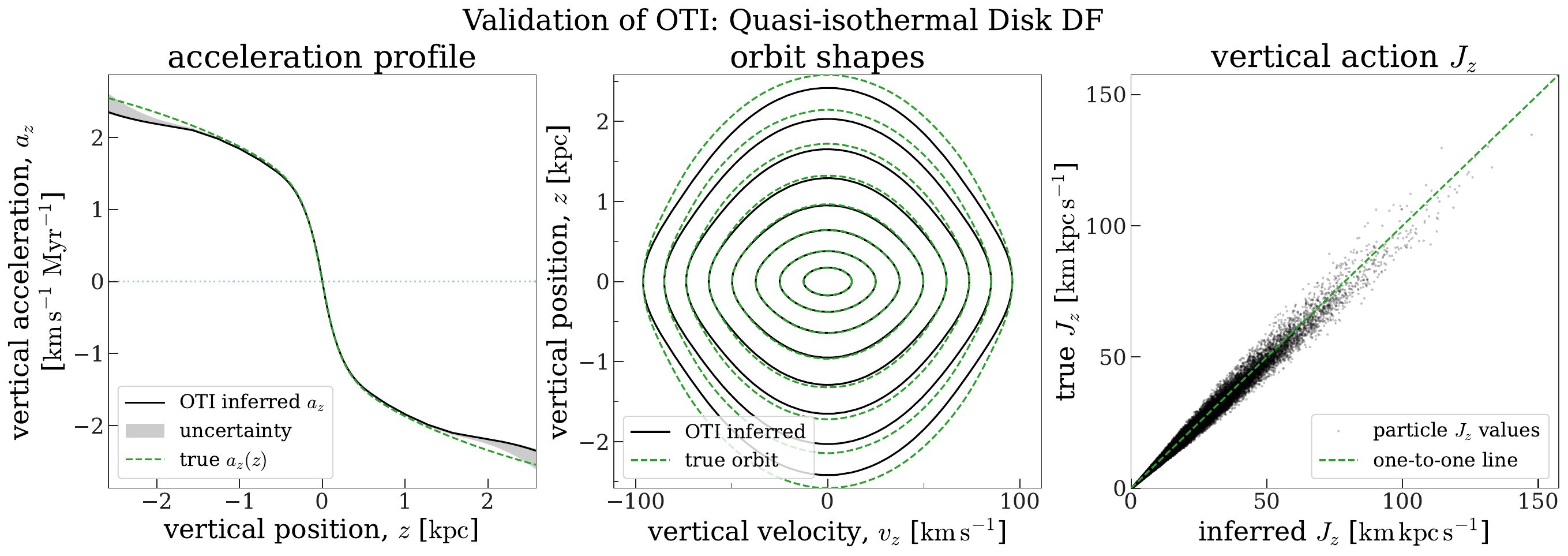}
\end{center}
\caption{%
Validation of the OTI model fit to the simulated quasi-isothermal disk sample (i.e. the
same as Figure~\ref{fig:sho-validation}, but for the simulated disk sample from
Section~\ref{sec:sim-qiso}).
The median (fractional) action error is $\approx 6\%$ for this sample dataset, but we do
not need to assume a gravitational potential to compute the actions.
\label{fig:qiso-validation}
}
\end{figure*}

We again use the optimized parameter values to initialize an MCMC sampling of the
parameters following the same procedure as in Section~\ref{sec:sim-sho}.
The left panel of Figure~\ref{fig:qiso-validation} shows the inferred vertical
acceleration profile (black line; computed with the MAP parameter values) and
uncertainties (gray shaded region; computed using the 16th--84th percentile parameter
values using the MCMC samples) along with the true vertical acceleration (green dashed
line) evaluated at a constant $R=R_0$ in our adopted Milky Way potential model (see
Appendix~\ref{sec:appendix-potential}).
The acceleration precision at $z=1~\kpc$ is $\sim 1\%$ (in fractional error) with a
$\sim 1\%$ bias.

The middle panel of Figure~\ref{fig:qiso-validation} shows eight orbits with
equally-spaced (but arbitrary) values of the vertical velocity at $z=0$: The
under-plotted solid (black) lines show the orbital trajectories inferred using the MAP
parameter values with our OTI model, and the over-plotted dashed (green) lines show the
true orbital trajectories.
The true orbital trajectories are computed in the 3D galaxy model using the grid of
$v_z$ values at $z=0$ with $x=R_0$, $v_x=0$, and $y=0$.
To set $v_y$ for the true orbits, we take the true actions computed for the particles in
our selected region and fit a 3rd-order polynomial to the relationship between $J_\phi$
and $J_z$ in this region (which arises primarily because of the finite selection on
$R$).
For the grid of $v_z$ values, we transform to $J_z$ and evaluate the fitted polynomial
to obtain a $J_\phi$ value, which we then convert to $v_y$ by dividing out the radius
$R_0$.
There is good agreement between the true and inferred acceleration profile and orbits,
especially for $|z| \lesssim 1~\kpc$.
Above $|z| \gtrsim 1~\kpc$, the true orbit shapes and the OTI-inferred orbits begin to
defer as the assumption of $R$--$z$ separability becomes less and less applicable.

We again compute the actions, frequencies, and angles for the star particles using
Equations~\ref{eq:Jz-ell}--\ref{eq:thz-ell} evaluated at the phase-space coordinates of
star particles in this simulated data set.
The right panel of Figure~\ref{fig:qiso-validation} shows (as black markers) the true
vertical action values for a random subset of $10^5$ simulated particles on the vertical
axis and our estimates for the vertical action of each star particle using the OTI
model on the horizontal axis.
We find the median fractional error of the vertical action, angle, and frequency are
$\sim 5\%$, $\sim 6\%$, and $\sim 10\%$, respectively.
But again, we do not need to assume a gravitational potential to compute the values.

\subsection{Axisymmetric Quasi-isothermal Disk with Survey Selection}
\label{sec:sim-qiso-sel}

Real data from stellar surveys often has complex selection effects resulting from survey
targeting and design, instrument limitations (e.g., magnitude limits), and/or real
astronomical effects like bright stars, crowding, or dust extinction.
All of these predominantly depend on position and brightness of the sources considered
and not strongly on derived parameters like element abundances.
One of the benefits of OTI is that positional selection effects do not bias our
inferences so long as there are no strong correlations between the effective selection
functions in position and the stellar labels considered.
Here we demonstrate this by repeating the experiment in Section~\ref{sec:sim-qiso}, but
now after applying a $z$-dependent selection function such that the probability a source
with a given $z$ value appears in our dataset $S(z)$ scales with
\begin{equation}
    S(z) \propto \left(\frac{|z| + 150~\unit{pc}}{2~\kpc}\right)^2 \quad .
\end{equation}

\begin{figure*}[t!]
\begin{center}
\includegraphics[width=\textwidth]{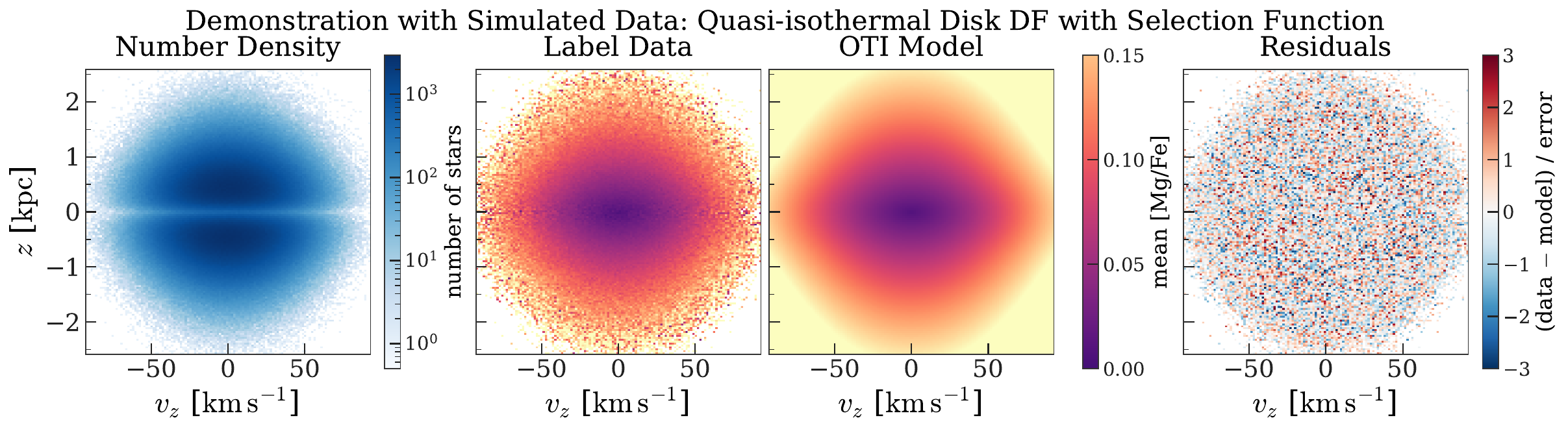}
\end{center}
\caption{%
The same as Figure~\ref{fig:sho-data-model}, but for the simulated quasi-isothermal disk
sample with an imposed selection function that depends on $z$ from
Section~\ref{sec:sim-qiso}.
The left two panels show the simulated data set, the middle right panel shows the
optimized OTI model, and the rightmost panel shows the residuals.
\label{fig:qiso-sel-data-model}
}
\end{figure*}

\begin{figure*}[t!]
\begin{center}
\includegraphics[width=\textwidth]{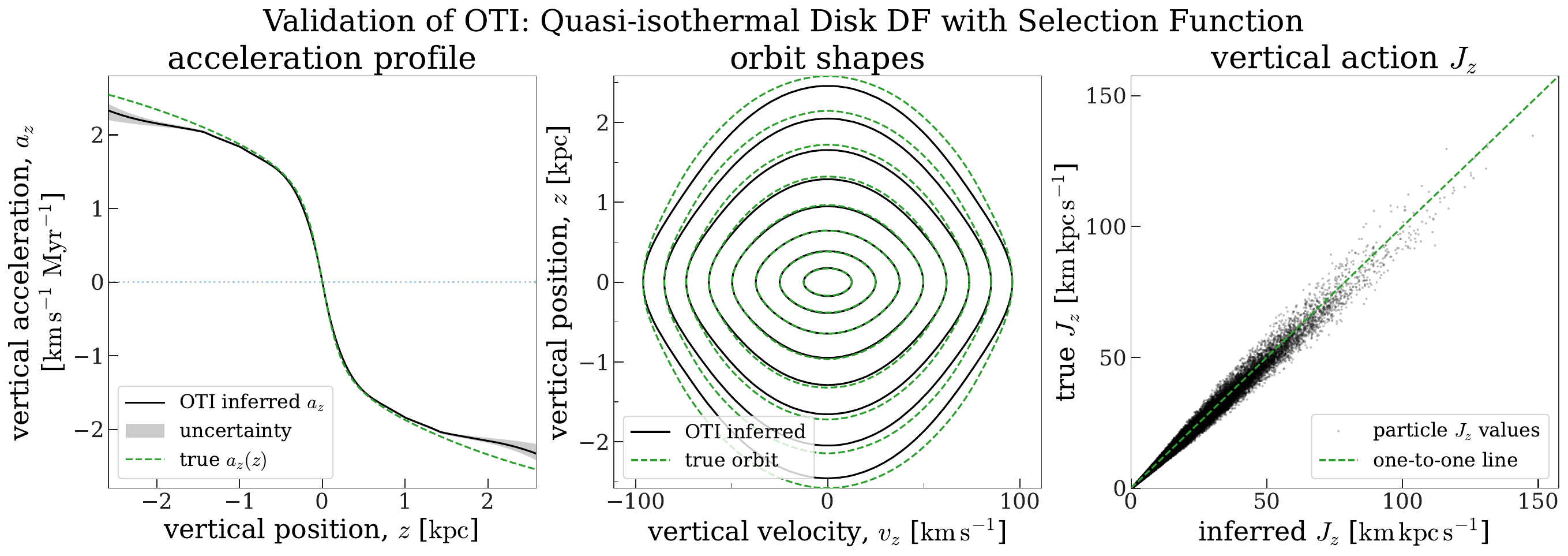}
\end{center}
\caption{%
Validation of the OTI model fit to the simulated quasi-isothermal disk sample with an
imposed selection function (i.e. the same as Figure~\ref{fig:sho-validation}, but for
the simulated disk sample from Section~\ref{sec:sim-qiso-sel}).
The model residuals in Figure~\ref{fig:qiso-sel-data-model} look reasonable, but this
model has an unphysical gradient of the vertical acceleration (which would lead to
negative density).
\label{fig:qiso-sel-validation}
}
\end{figure*}

Figure~\ref{fig:qiso-sel-data-model} again shows the input data, where the left two
panels show the number counts of stars per pixel of vertical phase-space coordinates and
the mean (simulated) \mgfe\ abundance per pixel.
The missing and lower density of stars near the midplane (mainly visible in the leftmost
panel, relative to Figure~\ref{fig:qiso-data-model}) is the effect of our simulated
selection function.
The middle right panel again shows the OTI model evaluated at the MAP parameters after
MCMC sampling, and the rightmost panel shows the residuals between the data and the MAP
model.
Even with the strong selection function, the OTI model leverages the strong (assumed)
symmetries in the phase space to infer the orbit shapes even when data are missing.
Figure~\ref{fig:qiso-sel-validation} shows the validation of the OTI model fit to the
data with a positional selection function imposed, demonstrating that the OTI model is
still able to recover the true vertical acceleration profile and orbital shapes where
our assumption of $R$--$z$ separability is most valid ($|z| \lesssim 1~\kpc$).

% Here, we see a potential pitfall of the flexibility of the OTI framework.
% For this and previous examples, we have used unconstrained monotonic spline functions to
% represent the label function and Fourier coefficient functions.
% This allows the model to fit the data in phase space extremely well, but can lead to
% unphysical behavior when interpreting the model in the context of dynamical quantities
% that we may want to infer.
% In this case, the gradient of the vertical acceleration changes sign at $|z| \approx
% 1~\kpc$, and there is no obvious reason why this happens (i.e. the optimizer completes
% successfully and the optimized model is a good representation of the phase space data).
% The fact that the gradient of the acceleration goes positive should not be allowed in a
% physical system, as the negative of the gradient of the acceleration is related to the
% density, which must always be positive.

% In this case, we are not concerned by the unphysical behavior at large $z$ because we
% know that our assumption of separability in $R$--$z$ is not valid at large $z$, so any
% results for $|z| \gtrsim 1~\kpc$ or so will, in general, be biased.
% We discuss this further in Section~\ref{sec:disc-tradeoff}, including some suggestions
% for how to mitigate failures due to the flexibility of the spline model.

\subsection{N-body Simulation with a Perturbed Disk}
\label{sec:sim-jason}

As a final demonstration with simulated data, we apply the OTI framework to a section of
an $N$-body simulation of a stellar disk that has been perturbed by an orbiting
satellite galaxy.
For this, we use the $N$-body simulation M1 of the \texttt{SMUDGE}\ simulation suite
\cite[]{Hunt:2021} now available on
\texttt{SciServer}\footnote{https://sciserver.org/datasets/smudge/} \citep{sciserver}.
M1 simulates the merger of a $\sim8\times10^{10}~\unit{\Msun}$ dwarf galaxy consisting
of two Hernquist spheres \citep{Hernquist:1990} into a $\sim6\times10^{11}~\unit{\Msun}$
disk galaxy host.
The host galaxy consists of a $8.8\times10^8$ particle live NFW halo
\citep{Navarro:1997}, a $2.2\times10^7$ particle Hernquist bulge and a $2.2\times10^8$
particle exponential disk with a high Toomre parameter $\mathcal{Q}=2.2$
\citep{Toomre:1964}, following model MWb from \cite{Widrow:2005}.
For full details of the initial condition and simulation setup see \cite{Hunt:2021}.

The merger simulation is evolved for $\sim8.3~\unit{\giga\year}$ using the GPU-based
$N$-body tree code \texttt{Bonsai}\ \citep{Bedorf:2012,Bedorf:2014}.
Following \cite{Hunt:2021} we choose snapshot 702 (occurring at $t=6.874$) as the
`present day' snapshot used in this work, as this is the snapshot where satellite is
closest to the location of the Sagittarius dwarf galaxy in the Milky Way.
However, we stress that model M1 is not meant to reproduce the Milky Way--Sagittarius
interaction \citep[for which the reader should see][]{Bennett:2022} but instead be a
laboratory for exploring satellite induced perturbation in an otherwise stable disk.
We use a snapshot of the simulation after four pericentric passages, and six disk
crossings of the satellite \citep[see Figure 3 of][for the orbit and mass loss of the
satellite]{Hunt:2021}, which induces significant non-axisymmetric kinematic structure in
the disk.
% Figure~\ref{fig:jason-mean-z-vz} shows the mean vertical position (left panel) and
% vertical velocity (right panel) of stars across the face of the simulated disk, showing
% that the disk is rich with substructure as a result of repeat satellite perturbations.
This application is a test of how well OTI can do (i.e. how biased the results are) in
the presence of strong disequilibrium, which violates the assumptions laid out above.

We select a small region of the disk near the solar radius, $R=8~\kpc$, by selecting all
star particles within $|x - -\kpc| < 1~\kpc$ and $|y| < 1~\kpc$.
We again mimic dynamical selections we could perform on real data by selecting stars
near their guiding-center radius $R_G$, i.e. $|R-R_G| < 1~\kpc$, and with small radial
velocity $|v_R| < 15~\kms$.
These dynamical selections limit the radial action of the stars so that our assumption
of $R$--$z$ separability is less invalid, but does not require computing actions for the
star particles before selection.
We compute the guiding radii using the circular velocity curve of the simulated disk,
$R_G \approx L_z / v_c(R)$, but here we use actions only to select a sample of stars on
nearly circular orbits with which to measure the circular velocity.
We compute the circular velocity by selecting star particles with low radial and
vertical action (below the 15th percentile values of the radial and vertical actions,
respectively) and compute the mean azimuthal velocity $\langle v_\phi(R) \rangle$ as a
function of cylindrical radius $R$, then adopt $v_c(R) = \langle v_\phi(R) \rangle$.

The ``true'' comparison actions are computed using \texttt{Agama}\
\citep{Vasiliev:2019}.
We first approximate the host galaxy potential using two multipole expansions for the
stellar bulge and dark matter halo, and an axisymmetric \texttt{CylSpline}\ expansion
for the disk.
We calculate actions, angles, and frequencies in the reconstructed potential using
\texttt{Agama}'s \texttt{ActionFinder}.
We note that, in this case, there is no ground truth for the actions, as the galaxy
potential model is only approximate and any method for computing actions will have some
additional approximation error.

We use these orbital actions to ``paint'' element abundances onto star particles in
this idealized $N$-body simulation without gas or star formation.
We compute the actions at an early snapshot of the simulation, after relaxation of the
initial conditions and before the first pericentric passage of the satellite galaxy.
We use the vertical actions of the star particles at this early snapshot to assign a
simulated \abun{Mg}{Fe} abundance value to each star particle following the procedure
explained in Section~\ref{sec:sim-qiso}.
We then track the abundance values through to the final snapshot where the actions have
evolved due to satellite perturbations.
To better emphasize the disequilibrium in the space of abundances, we reduce the
intrinsic scatter of the abundances relative to the value estimated from \apogee\ data
(Figure~\ref{fig:mgfe-zvz}) to be $0.05 / 4$ instead of $0.05$.

\begin{figure*}[t!]
\begin{center}
\includegraphics[width=\textwidth]{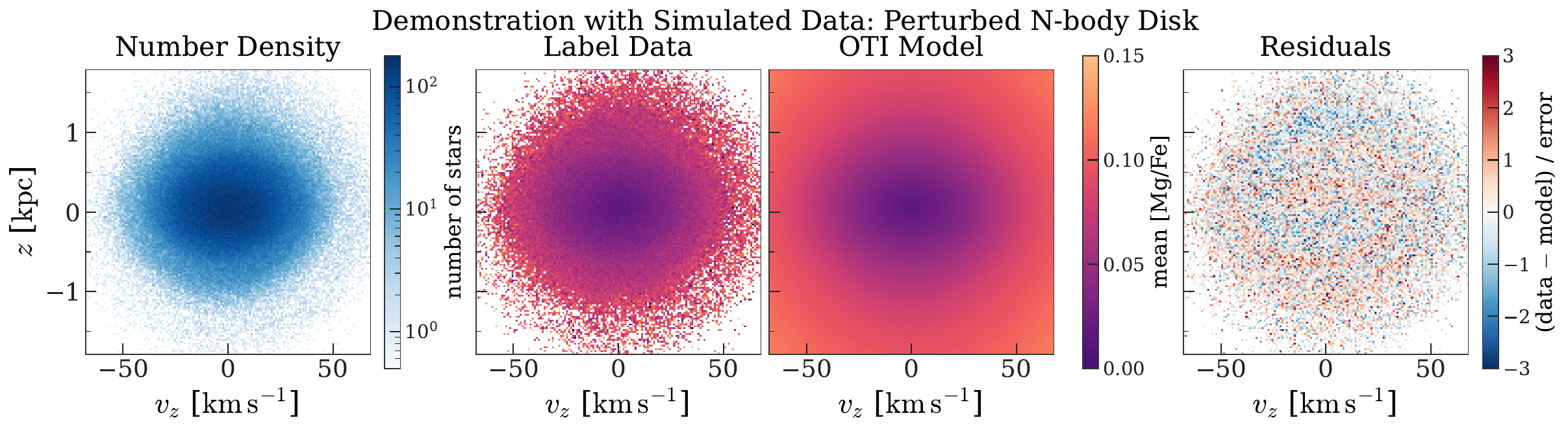}
\end{center}
\caption{%
The same as Figure~\ref{fig:sho-data-model}, but for data from a small region of a live
$N$-body simulation of a disk galaxy perturbed by an orbiting satellite on an orbit
similar to that predicted for the Sagittarius dwarf galaxy.
The vertical structure of the simulated disk is perturbed repeatedly by interactions
with the satellite, which create a vertical phase spiral in the number density of stars
(leftmost panel) and weaker signatures of disequilibrium in simulated element abundances
(middle left panel).
The middle right panel again shows the optimized OTI model for these data, and the
rightmost panel shows the (normalized) residuals.
\label{fig:jason-data-model}
}
\end{figure*}

The left two panels of Figure~\ref{fig:jason-data-model} show the number density of
stars in this volume, and the mean simulated \abun{Mg}{Fe} abundance of star particles
in the vertical phase space of the ``solar neighborhood'' of the simulated stellar disk.
While the overall pattern is similar to the smooth examples (e.g.,
Section~\ref{sec:sim-qiso}), the mean abundance values here show appreciable asymmetries
due to the perturbations from the simulated satellite galaxy.
The middle right panel of Figure~\ref{fig:jason-data-model} shows the mean \abun{Mg}{Fe}
of the optimized OTI model for this simulated data set, and the rightmost panel again
shows the residuals.
While the residuals are generally low, the residuals here have clear structure: There is
a faint spiral pattern in the abundance residuals that is an analog to the vertical
phase spiral seen in \gaia\ data in the number densities of stars.
The presence of this abundance spiral in this simulated data should bias our inferred
dynamical quantities.

\begin{figure*}[t!]
\begin{center}
\includegraphics[width=\textwidth]{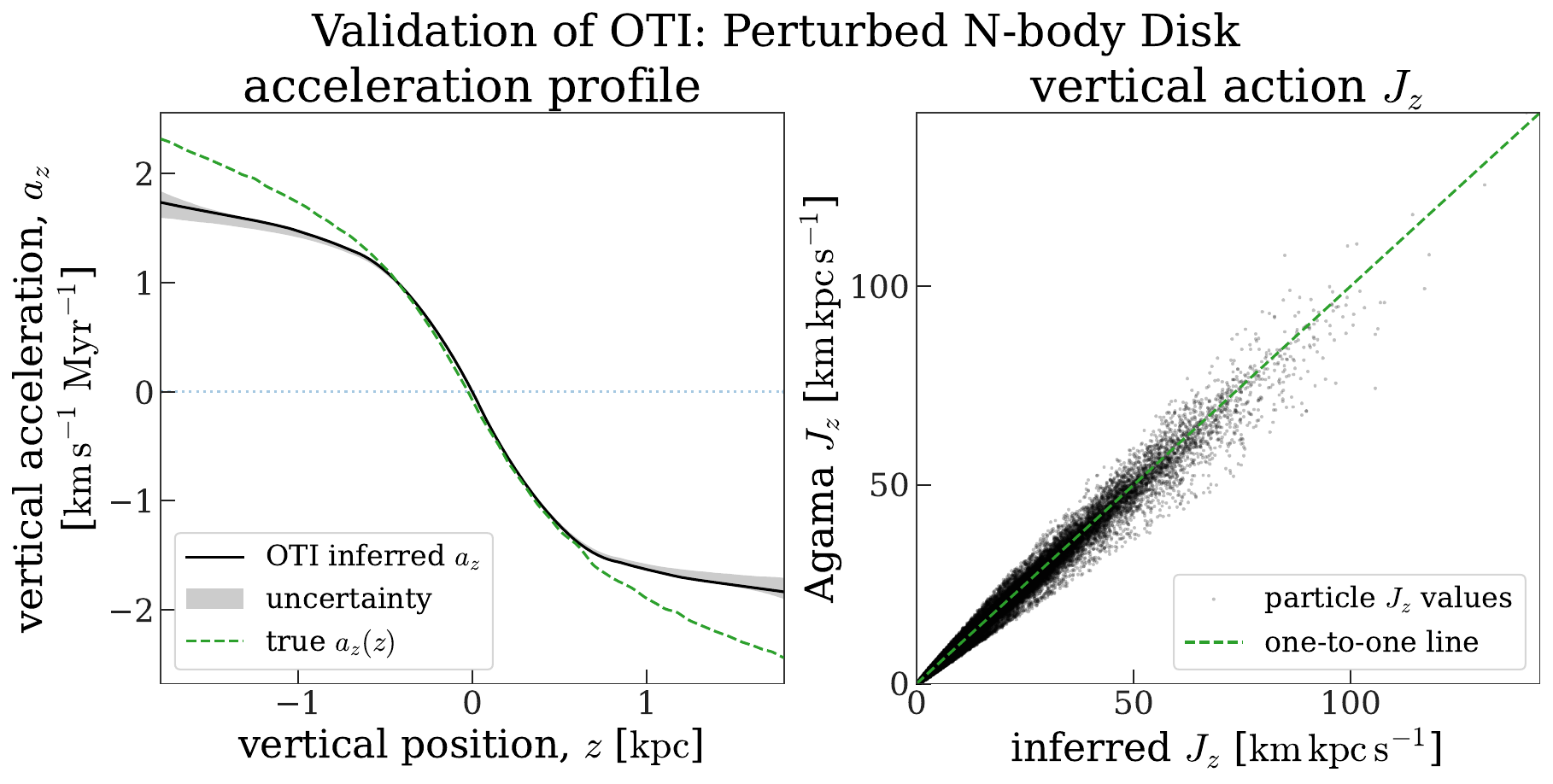}
\end{center}
\caption{%
Validation of the OTI model fit to the vertical phase space of a small volume from an
$N$-body simulation of a disk galaxy perturbed by an orbiting satellite.
The left panel shows the inferred vertical acceleration profile and uncertainties (black
line and gray shaded region) compared to the true acceleration profile (green dashed
line).
The inferred acceleration profile is biased for $|z| \gtrsim 0.5~\kpc$ by the
disequilibrium.
The right panel shows a comparison of vertical action values computed with the best-fit
OTI model (horizontal axis) and with Agama, using a time-independent reconstruction of
the total gravitational potential of the simulated disk and dark matter halo (vertical
axis).
\label{fig:jason-validation}
}
\end{figure*}

Figure~\ref{fig:jason-validation} (left panel) shows the vertical acceleration profile
inferred by OTI for this region of the simulated disk (black line and gray shaded
region) compared to the true acceleration profile (green dashed line).
Within $|z| \lesssim 0.5~\kpc$, the OTI model agrees well with the true acceleration
profile.
However, for $|z| \gtrsim 0.5~\kpc$ the OTI model clearly diverges from the truth with a
bias at $|z| = 1~\kpc$ of about $15\%$.
The right panel of Figure~\ref{fig:jason-validation} shows a comparison of vertical
action values computed with the best-fit OTI model (horizontal axis) and with Agama
(vertical axis).
As mentioned above, the actions computed with Agama are not necessarily ``true'' values,
as this system is not in equilibrium or symmetric, but the adopted potential model used
to compute the actions is both axisymmetric and time-invariant.
We find a median fractional difference of $\sim 6\%$ for the vertical action values,
$\sim 12\%$ for the vertical frequencies, and $\sim 5\%$ for the vertical angles.
For real Milky Way stars, we expect similar biases due to the presence of the vertical
phase spiral \citep{Antoja:2018}.

% Note: moved phase spiral / Gaia DR3 stuff to adrn/oti-phase-spiral

\section{Discussion} \label{sec:discussion}

In the subsections below, we return to our assumptions and several of the points raised
in the results above to discuss the implications of our work and the limitations of the
Orbital Torus Imaging (OTI) framework.

\subsection{The Tradeoff of Flexibility and Physicality} \label{sec:disc-tradeoff}

The OTI framework provides a flexible means for inferring the orbit structure and,
therefore, the mass distribution underlying phase-mixed, tracer stellar populations.
It works by parameterizing and modeling the shapes of contours of constant stellar
labels in phase space.
By assuming that these contours correspond to orbits, we can use the changing shapes of
the contours to infer the underlying total gravitational acceleration profile traced by
the observed stars.
In this way, it decouples the problem of modeling the phase space variation of stellar
labels from the interpretation step, when this variation is mapped to the underlying
gravitational field.
It is flexible in that it does not require a global model for the gravitational
potential or the \df, but instead uses the data itself to infer the orbit structure.
However, this flexibility and this separation of modeling the phase space from physical
interpretation can come at the cost of physicality.

Though it is possible to instantiate an OTI model with specific physical constraints
(e.g., that the orbits follow a gravitational potential and the density is positive),
the general OTI framework does not require this.
As was shown with our demonstrations above (Section~\ref{sec:applications-sim}), this
can lead to unphysical inferred acceleration profiles that arise both because of the
presence of disequilibrium and because of the breakdown of our assumption of $R$--$z$
separability (Sections~\ref{sec:sim-qiso-sel}--\ref{sec:sim-jason}).
What we gain from this flexibility is the ability to fit the phase space data very well,
in ways that more constrained equilibrium models may find stronger residuals.
Depending on the scientific goal, this may be a worthwhile tradeoff.
For example, one may only want a flexible, symmetric model of the phase space to act as
a filter to subtract and study the morphology of substructure (e.g., the \gaia\ phase
spiral; \citealt{Antoja:2018}).
If one wants to infer the acceleration profile or other dynamical quantities (e.g.,
orbital actions, frequencies, or angles), the model must be validated by assessing the
physicality of the inferred quantities.
Or, one must use a more constrained model by adopting functional forms for the Fourier
distortion functions and/or the label function that only permit physical solutions for
the acceleration and density field.
We leave this as a topic for future work.

\subsection{The Impact of Disequilibrium} \label{sec:disc-diseq}

The Milky Way's disk is now known to be out of equilibrium, especially in terms of the
vertical kinematics of stars \citep[e.g.,][]{Antoja:2018, Katz:2018, Ramos:2018,
Tian:2018, Laporte:2019, Khanna:2019, Hunt:2022, Darragh-Ford:2023}.
The imprint of the vertical ``phase spiral'' is apparent in both the phase-space density
in the Milky Way and even in the mean element abundances of stars
\citep{Frankel:inprep}.
Non-phase-mixed structures in the vertical kinematics violate our assumptions and will
generally bias our inferred orbit shapes and acceleration profile, as demonstrated in
Section~\ref{sec:sim-jason}.
In this work, we simply highlight this potential bias and do not attempt to correct for
this in our demonstrations above.
However, there is a path toward accounting for weak disequilibrium in the framework
outlined above.

In the limit that the dependence of stellar label moments on the phase-space coordinates
is a sum of a smooth, equilibrium component and a low-amplitude, phase-coherent
component, we can build the phase-dependence of the stellar labels directly into the OTI
model.
In what we have done in this work, in terms of the vertical phase space, we have assumed
that the stellar labels $Y$ or moments of the stellar labels $\langle Y \rangle$ only
depend on the distorted elliptical radius $\rz$ (Equation~\ref{eq:rz}) such that, e.g.,
$Y = Y(\rz)$.
We can generalize this to allow for a phase-coherent component by instead making $Y =
Y(\rz, \theta_z)$, where $\theta_z$ is the vertical angle (Equation~\ref{eq:thz-ell}),
which can be computed internally within the OTI model for a given setting of the
parameters.
We consider a full exploration of this extensions as out of scope for this work, but
hope to explore this in the future.

\subsection{Beyond Vertical Dynamics} \label{sec:disc-beyond-vertical}

While this work primarily focuses on the vertical dynamics of stars in the Milky Way
disk, a natural extension of the current description of the OTI framework is to consider
the radial $R$ and azimuthal $\phi$ dependencies as well.
In principle, the OTI framework is applicable in any projection of phase space, meaning
it is possible to extend the framework to model the dynamics in the entire 6D phase
space.
Incorporating the radial dependence of the stellar labels and their moments would allow
us to connect and jointly model regions of the disk in a flexible framework that does
not require imposing a global model of the potential or density distribution.
As it is currently implemented, we can only achieve this by binning the data into
regions of $R$ and/or $\phi$ to consider a single projection of the phase space.
However, even within this model, we could explore instead the $\Delta R = R - R_G$ vs.
$v_R$ phase space as a complementary view of the dynamics of the Milky Way disk. We hope
to explore this in future work.

Another avenue for exploration is the possibility of going coordinate-free, or at least
projection-free, to identify surfaces in the full 6D phase space upon which stellar
label moments remain approximately constant.
Under the assumptions described above, these surfaces would correspond to the orbital
tori of stars and would allow us to directly map the Galaxy's orbit structure.
A first exploration of this more general path is considered in \cite{Novara:inprep}.

\subsection{What stellar labels are useful for OTI?} \label{sec:disc-what-labels}

As mentioned above (Section~\ref{sec:sim-qiso-sel}), a critical requirement for OTI is
that the selection function on stellar phase-space coordinates is separable from the
selection function of the stellar labels considered (see Section~\ref{sec:intro}).
For element abundances, this is mostly true for surveys like \apogee, where the primary
survey target selection is done based on color and magnitude, which only weakly depend
on surface abundances \citep{Zasowski:2017, Santana:2021, Beaton:2021}.
However, this also implies that stellar surface gravity, $\logg$, or effective
temperature, $\Teff$, would \emph{not} be useful stellar labels within the OTI framework
because they are strongly covariant with luminosity or absolute magnitude.
Additionally, there is a potential danger in using kinematic measurements that are
inferred from spectroscopy (e.g., spectrophotometric distances; \citealt{Hogg:2019,
Leung:2019}) in OTI:
If the kinematic measurements are inferred from the spectra, even if the target
selection is separable, the estimators may inject correlations between the stellar
labels and the phase-space coordinates.

The method presented here works for any moments of any stellar labels that satisfy the
condition that the selection function on the stellar labels is separable from the
selection function on the phase-space coordinates.
Can we therefore combine the likelihoods from multiple element abundances or multiple
moments of the abundances to improve the precision of our inferences?
In principle, yes, but element abundances tend to be strongly correlated with each other
\citep[e.g.,][]{Ness:2019, Griffith:2023}, so the different moments and/or abundances
must be incorporated conditionally \citep[e.g.,][]{Ratcliffe:2023}.
We will explore this in future work.

\section{Summary and Conclusions} \label{sec:conclusions}

We introduce a new formulation and implementation of the Orbital Torus Imaging (OTI)
framework.
We derive a rigorous justification for OTI based on dynamical theory, and demonstrate
its utility for inferring the orbital structure of stellar populations and
characterizing the underlying gravitational field using stellar labels.
OTI works by parameterizing and modeling the shapes of contours of constant stellar
labels or constant moments of the stellar label distribution in regions of phase space.
By assuming that these contours correspond to orbits, we use the changing shapes of the
contours to infer the underlying total gravitational acceleration profile traced by the
observed stars.
OTI is less sensitive to selection effects than traditional methods that model the
stellar phase-space density, and does not require specifying a global model for the
gravitational potential or the stellar distribution function.
We demonstrate the flexibility of OTI in modeling and recovering the true underlying
gravitational field for simulated tracer stellar populations in toy and more realistic
settings in the vertical phase space.
However, the mathematical formulation of OTI applies to a wider range of systems, both
with higher dimensions (e.g., axisymmetric systems) and out of equilibrium (e.g., with
weak disequilibrium).
The OTI framework is implemented in the open-source \texttt{Python} package
\texttt{torusimaging} \citep{torusimaging:zenodo}.
In our companion paper, we use the OTI framework to measure the vertical acceleration
profile and the mass distribution of the Milky Way disk using \apogee\ data
\citep{Horta:2023}.
Future data releases from \gaia, \sdssv, and upcoming spectroscopic surveys will enable
new and more precise measurements of the Milky Way's gravitational field using OTI.

\begin{acknowledgements}

It is a pleasure to thank Neige Frankel, Melissa Ness, Julianne Dalcanton, and the
\textit{Nearby Universe Group} at the CCA for useful discussions and feedback on this
work.

KVJ's contributions were supported by a grant from the Simons Foundation (CCA 1018465).
LMW was supported by a Discovery Grant with the Natural Sciences and and Engineering
Research Council of Canada.

Funding for the Sloan Digital Sky Survey IV has been provided by the Alfred P.
Sloan Foundation, the U.S. Department of Energy Office of Science, and the
Participating Institutions. SDSS-IV acknowledges support and resources from the
Center for High-Performance Computing at the University of Utah. The SDSS web
site is www.sdss.org.

SDSS-IV is managed by the Astrophysical Research Consortium for the
Participating Institutions of the SDSS Collaboration including the Brazilian
Participation Group, the Carnegie Institution for Science, Carnegie Mellon
University, the Chilean Participation Group, the French Participation Group,
Harvard-Smithsonian Center for Astrophysics, Instituto de Astrof\'isica de
Canarias, The Johns Hopkins University, Kavli Institute for the Physics and
Mathematics of the Universe (IPMU) / University of Tokyo, Lawrence Berkeley
National Laboratory, Leibniz Institut f\"ur Astrophysik Potsdam (AIP),
Max-Planck-Institut f\"ur Astronomie (MPIA Heidelberg), Max-Planck-Institut
f\"ur Astrophysik (MPA Garching), Max-Planck-Institut f\"ur Extraterrestrische
Physik (MPE), National Astronomical Observatories of China, New Mexico State
University, New York University, University of Notre Dame, Observat\'ario
Nacional / MCTI, The Ohio State University, Pennsylvania State University,
Shanghai Astronomical Observatory, United Kingdom Participation Group,
Universidad Nacional Aut\'onoma de M\'exico, University of Arizona, University
of Colorado Boulder, University of Oxford, University of Portsmouth, University
of Utah, University of Virginia, University of Washington, University of
Wisconsin, Vanderbilt University, and Yale University.

This work has made use of data from the European Space Agency (ESA) mission
{\it Gaia} (\url{https://www.cosmos.esa.int/gaia}), processed by the {\it Gaia}
Data Processing and Analysis Consortium (DPAC,
\url{https://www.cosmos.esa.int/web/gaia/dpac/consortium}). Funding for the DPAC
has been provided by national institutions, in particular the institutions
participating in the {\it Gaia} Multilateral Agreement.

\end{acknowledgements}

\software{
    Agama \citep{Vasiliev:2019},
    Astropy \citep{astropy:2013, astropy:2018, astropy:2022},
    gala \citep{gala},
    % IPython \citep{ipython},
    JAX \citep{jax:2018},
    JAXOpt \citep{jaxopt:2021},
    matplotlib \citep{Hunter:2007},
    numpy \citep{numpy},
    pyia \citep{pyia},
    scipy \citep{scipy},
    torusimaging \citep{torusimaging:zenodo}.
}

\appendix

\section{A Toy Milky Way Mass Model}
\label{sec:appendix-potential}

For the demonstrations with equilibrium models above
(Section~\ref{sec:sim-qiso}--\ref{sec:sim-qiso-sel}), we use a simplistic mass model to
represent the total gravitational potential of the Milky Way.
We use a two component model consisting of a Miyamoto--Nagai disk \citep{Miyamoto:1975}
embedded in a spherical Navarro--Frenk--White (NFW) halo \citep{Navarro:1997}.
We adopt the following parameter values for the disk and halo, respectively: $M_{\rm
disk} = 6.91\times 10^{10}~\unit{\Msun}$, $a = 3~\kpc$, $b = 0.25~\kpc$, $M_{\rm halo} =
5.4\times 10^{11}~\unit{\Msun}$, $r_s = 15~\kpc$, where $a$ and $b$ are the disk scale
length and scale height, respectively, $M_{\rm halo}$ is the scale mass of the halo (not
the virial mass), and $r_s$ is the scale radius of the halo.
These values are chosen so that the circular velocity at the solar radius $R_0 =
8.3~\kpc$ is $v_c(R_0) = 229~\kms$, and other parameters are chosen to be consistent
with the \texttt{MilkyWayPotential2022} implemented in \texttt{gala} \citep{gala}.

% \section{Monotonic Quadratic Spline}
% \label{sec:appendix-spline}

% TODO

\section{Evaluation of the Vertical Acceleration from an OTI Model}
\label{sec:appendix-az}

We evaluate the vertical acceleration as a function of height above the Galactic
midplane, $a_z(z) = \deriv{\Phi}{z}$, given an OTI model and parameters using
Equation~\ref{eq:az-rz}, reproduced here:
\begin{equation}
    a_z = - v_z \, \frac{\partial \rz}{\partial z} \,
        \left( \frac{\partial \rz}{\partial v_z} \right)^{-1}
\end{equation}
The acceleration only depends on the shapes of the contours of the DF, i.e. curves of
constant \rz, so that we only need to differentiate the distorted radius to evaluate it.
Based on our definitions above (Equations~\ref{eq:rzp}, \ref{eq:rz}, and
\ref{eq:thz-ell}), the partial derivatives of \rz\ with respect to $z$ and $v_z$ are:
\begin{align}
    \frac{\partial \rz}{\partial z} &=
        \frac{\partial \rzp}{\partial z} +
        \sum_{m=2}^{\mmax} \frac{\partial}{\partial z} \left[\rzp \, e_m(\rzp) \, \cos\left(m \, \thzp\right)\right]\\
    \frac{\partial \rz}{\partial v_z} &=
        \frac{\partial \rzp}{\partial v_z} +
        \sum_{m=2}^{\mmax} \frac{\partial}{\partial v_z} \left[\rzp \, e_m(\rzp) \, \cos\left(m \, \thzp\right)\right]
\end{align}
Plugging these in to Equation~\ref{eq:az-rz}, and using the following expressions
\begin{alignat}{3}
    \frac{\partial \rzp}{\partial z} &= \frac{z\,\freqzero}{\rzp} \quad &; \quad
        \frac{\partial \rzp}{\partial v_z} &= \frac{v_z}{\freqzero \, \rzp} \\
    \frac{\partial \thzp}{\partial z} &= \frac{v_z}{\rzp^2} \quad &; \quad
        \frac{\partial \thzp}{\partial v_z} &= -\frac{z}{\rzp^2} \\
    \frac{\partial e_m}{\partial z} &=
        \frac{\partial e_m}{\partial \rzp} \, \frac{\partial \rzp}{\partial z}
        \quad &; \quad
        \frac{\partial e_m}{\partial v_z} &=
        \frac{\partial e_m}{\partial \rzp} \, \frac{\partial \rzp}{\partial v_z}
\end{alignat}
our goal is to remove or collect terms that explicitly contain $v_z$, as the
acceleration should not depend on velocity (though we do not require this in the OTI
framework).
With this in mind, and using the expressions listed above to simplify, we find that
\begin{align}
    \frac{\partial \rz}{\partial z} &=
        \frac{\partial \rzp}{\partial z} +
        \sum_{m=2}^{\mmax} \frac{\partial}{\partial z}
            \left[\rzp \, e_m(\rzp) \, \cos(m \, \thzp)\right]\\
    % &= \frac{\partial \rzp}{\partial z} +
    %     \sum_{m=2}^{\mmax} \left[
    %     \frac{\partial \rzp}{\partial z} \, e_m \, \cos(m \, \thzp)
    %     + \rzp \, \frac{\partial e_m}{\partial z} \, \cos(m \, \thzp)
    %     - \rzp \, e_m \, \sin(m \, \thzp) \, m \,\frac{\partial \thzp}{\partial z}\right]\\
    % &= \frac{z\,\freqzero}{\rzp} + \sum_{m=2}^{\mmax} \left[
    %     \frac{z\,\freqzero}{\rzp} \, e_m \, \cos(m \, \thzp) +
    %     \rzp \, \frac{z\,\freqzero}{\rzp} \, \frac{\partial e_m}{\partial \rzp} \, \cos(m \, \thzp) -
    %     \rzp \, e_m \, \sin(m \, \thzp) \, m \, \frac{v_z}{\rzp^2}\right]\\
    % &= \frac{z\,\freqzero}{\rzp} \, \left[
    %     1 + \sum_{m=2}^{\mmax} \left(
    %         e_m \, \cos(m \, \thzp) +
    %             \rzp \, \frac{\partial e_m}{\partial \rzp} \, \cos(m \, \thzp) -
    %             m \, e_m \, \sin(m \, \thzp) \, \frac{v_z}{z\,\freqzero}
    %     \right)
    % \right]\\
    &= \frac{z\,\freqzero}{\rzp} \, \left[
        1 + \sum_{m=2}^{\mmax} \left(
            \left(e_m + \rzp \, \deriv{e_m}{\rzp}\right) \,
                \cos(m \, \thzp) -
            m \, e_m \, \frac{\sin(m \, \thzp)}{\sin(\thzp)} \, \cos(\thzp)
        \right)
    \right] \label{eq:drz-pre-lim}
\end{align}
As we are interested in evaluating this expression along the $z$-axis, we will take the
limit of this expression as $\thzp \to \frac{\pi}{2}$, which is equivalent to the limit
$v_z \to 0$.
For even values of $m$, as we require, the sine term in the sum of Equation~\ref{eq:drz-pre-lim} will be zero, and the cosine terms become:
\begin{equation}
    \cos(m \, \thzp) = (-1)^{m/2} \quad .
\end{equation}
In this limit, the expression becomes:
\begin{equation}
    \lim_{\thzp \to \frac{\pi}{2}} \pderiv{\rz}{z} =
        \frac{z\,\freqzero}{\rzp} \, \left[
            1 + \sum_{m=2}^{\mmax} (-1)^{m/2} \,
                \left(e_m + \rzp \, \deriv{e_m}{\rzp}\right)
        \right] \quad .
\end{equation}

Similarly for the derivative with respect to $v_z$:
\begin{align}
    \pderiv{\rz}{v_z} &=
        \pderiv{\rzp}{v_z} +
        \sum_{m=2}^{\mmax} \frac{\partial}{\partial v_z}
            \left[\rzp \, e_m(\rzp) \, \cos(m \, \thzp)\right]\\
    % &= \frac{\partial \rzp}{\partial v_z} +
    %     \sum_{m=2}^{\mmax} \left[
    %     \frac{\partial \rzp}{\partial v_z} \, e_m \, \cos(m \, \thzp)
    %     + \rzp \, \frac{\partial e_m}{\partial v_z} \, \cos(m \, \thzp)
    %     - \rzp \, e_m \, \sin(m \, \thzp) \, m \,\frac{\partial \thzp}{\partial v_z}\right]\\
    % &= \frac{v_z}{\freqzero\,\rzp} + \sum_{m=2}^{\mmax} \left[
    %     \frac{v_z}{\freqzero\,\rzp} \, e_m \, \cos(m \, \thzp) +
    %     \rzp \, \frac{v_z}{\freqzero\,\rzp} \, \frac{\partial e_m}{\partial \rzp} \, \cos(m \, \thzp) +
    %     \rzp \, e_m \, \sin(m \, \thzp) \, m \, \frac{z}{\rzp^2}\right]\\
    % &= \frac{v_z}{\freqzero \, \rzp} \, \left[
    %     1 + \sum_{m=2}^{\mmax} \left(
    %         e_m \, \cos(m \, \thzp) +
    %             \rzp \, \frac{\partial e_m}{\partial \rzp} \, \cos(m \, \thzp) +
    %             m \, e_m \, \sin(m \, \thzp) \, \tan(\thzp)
    %     \right)
    % \right]\\
    &= \frac{v_z}{\freqzero \, \rzp} \, \left[
        1 + \sum_{m=2}^{\mmax} \left(
            \left(e_m + \rzp \, \deriv{e_m}{\rzp}\right) \,
                \cos(m \, \thzp) +
            m \, e_m \, \sin(m \, \thzp) \, \tan(\thzp)
        \right)
    \right]
\end{align}
Taking the same limit $\thzp \to \frac{\pi}{2}$, this expression becomes:
\begin{equation}
    \lim_{\thzp \to \frac{\pi}{2}} \frac{\partial \rz}{\partial v_z} =
        \frac{v_z}{\freqzero \, \rzp} \, \left[
            1 + \sum_{m=2}^{\mmax} (-1)^{m/2} \,
                \left(e_m\,(1 - m^2) + \rzp \, \deriv{e_m}{\rzp}\right)
        \right] \quad .
\end{equation}

Combining these expressions, we have:
\begin{align}
    a_z &=
        -v_z \, \frac{\partial \rz}{\partial z} \,
        \left( \frac{\partial \rz}{\partial v_z} \right)^{-1}\\
    &= -v_z \, \frac{z\,\freqzero}{\rzp} \, \frac{\freqzero \, \rzp}{v_z} \,
        \frac{\left[
            1 + \sum_{m=2}^{\mmax} (-1)^{m/2} \,
                \left(e_m + \rzp \, \deriv{e_m}{\rzp}\right)
        \right]}{\left[
            1 + \sum_{m=2}^{\mmax} (-1)^{m/2} \,
                \left(e_m\,(1 - m^2) + \rzp \, \deriv{e_m}{\rzp}\right)
        \right]} \\
    &= -\freqzero^2 \, z \,
        \frac{\left[
            1 + \sum_{m=2}^{\mmax} (-1)^{m/2} \,
                \left(e_m + \rzp \, \deriv{e_m}{\rzp}\right)
        \right]}{\left[
            1 + \sum_{m=2}^{\mmax} (-1)^{m/2} \,
                \left(e_m\,(1 - m^2) + \rzp \, \deriv{e_m}{\rzp}\right)
        \right]} \label{eq:az-after-lim}
\end{align}
This expression is evaluated at $\rzp = \sqrt{\freqzero} \, z$.
The multiplicative factor at the front of Equation~\ref{eq:az-after-lim} is the
acceleration profile of a simple harmonic oscillator, and the ratio of the sums captures
the changing orbit shapes due to the Fourier distortion functions.

\bibliographystyle{aasjournal}
\bibliography{refs}

\end{document}

%% file: preamble.tex
\newcommand{\package}[1]{\texttt{#1}}
\newcommand{\acronym}[1]{{\small{#1}}}

\newcommand{\python}{\textsl{Python}}
\newcommand{\jax}{\package{JAX}}
\newcommand{\agama}{\package{Agama}}

\newcommand{\given}{\,|\,}

\newcommand{\dd}{\mathrm{d}}
\newcommand{\deriv}[2]{\frac{\mathrm{d}{#1}}{\mathrm{d}{#2}}}

\newcommand{\Deriv}[2]{\frac{\mathrm{D}{#1}}{\mathrm{D}{#2}}}
\newcommand{\pderiv}[2]{\frac{\partial {#1}}{\partial {#2}}}

\usepackage{savesym}
\savesymbol{tablenum}
\usepackage{siunitx}
\restoresymbol{SIX}{tablenum}
\DeclareSIUnit\year{yr}
\DeclareSIUnit\parsec{pc}
\DeclareSIUnit\Msun{M_\odot}
\DeclareSIUnit\Rsun{R_\odot}

\newcommand{\kms}{\unit{\km\per\s}}
\newcommand{\kpc}{\unit{\kilo\parsec}}

\newcommand{\rzunit}{\unit{\kilo\parsec\per\mega\year\tothe{1/2}}}

\newcommand{\bs}[1]{\boldsymbol{#1}}

\newcommand{\abun}[2]{\ensuremath{{[\mathrm{#1}/\mathrm{#2}]}}}

\newcommand{\mgfe}{\abun{Mg}{Fe}}
\newcommand{\logg}{\ensuremath{\log g}}
\newcommand{\Teff}{\ensuremath{T_{\textrm{eff}}}}

\newcommand{\df}{\acronym{DF}}
\newcommand{\zmax}{\ensuremath{z_{\textrm{max}}}}

\newcommand{\gaia}{\textsl{Gaia}}
\newcommand{\dr}[1]{\acronym{DR}#1}
\newcommand{\apogee}{\acronym{APOGEE}}

\newcommand{\sdssv}{\acronym{SDSS-V}}